\documentclass[iop]{emulateapj}

\newcommand{\kms}{\rm km~s^{-1}}
\newcommand{\kmsmpc}{\rm km~s^{-1}~Mpc^{-1}}
\newcommand{\dn}{D_{n}4000}

\slugcomment{\bf Last updated:\today}

\usepackage{color}
\usepackage{multirow}
\usepackage{enumitem}

\begin{document}

\title{The Massively Accreting Cluster A2029} 

\author{Jubee Sohn$^{1}$, 
        Margaret J. Geller$^{1}$, Stephen A. Walker$^{2}$, Ian Dell'Antonio$^{3}$, 
        Antonaldo Diaferio$^{4, 5}$, Kenneth J. Rines$^{6}$}

\affil{$^{1}$ Smithsonian Astrophysical Observatory, 60 Garden Street, Cambridge, MA 02138, USA}
\affil{$^{2}$ Astrophysics Science Division, X-ray Astrophysics Laboratory, Code 662, NASA Goddard Space Flight Center, Greenbelt, MD 20771, USA}
\affil{$^{3}$ Department of Physics, Brown University, Box 1843, Providence, RI 02912, USA}
\affil{$^{4}$ Universit{\`a} di Torino, Dipartimento di Fisica, Torino, Italy}
\affil{$^{5}$ Istituto Nazionale di Fisica Nucleare (INFN), Sezione di Torino, Torino, Italy}
\affil{$^{6}$ Department of Physics and Astronomy, Western Washington University, Bellingham, WA 98225, USA}

\begin{abstract}
We explore the structure of galaxy cluster Abell 2029 and 
 its surroundings based on intensive spectroscopy along with X-ray and weak lensing observations. 
The redshift survey includes 4376 galaxies (1215 spectroscopic cluster members) within 40\arcmin  of the cluster center; 
 the redshifts are included here. 
Two subsystems, A2033 and a Southern Infalling Group (SIG) appear in the infall region 
 based on the spectroscopy as well as on the weak lensing and X-ray maps. 
The complete redshift survey of A2029 also identifies 
 at least 12 foreground and background systems (10 are extended X-ray sources) in the A2029 field;
 we include a census of their properties.
The X-ray luminosities ($L_{X}$) -- velocity dispersions ($\sigma_{cl}$) scaling relations
 for A2029, A2033, SIG, and the foreground/background systems 
 are consistent with the known cluster scaling relations.
The combined spectroscopy, weak lensing, and X-ray observations provide 
 a robust measure of the masses of A2029, A2033, and SIG. 
The total mass of the infalling groups (A2033 and SIG) is $\sim 60\%$ of the M$_{200}$ of the primary cluster, A2029.  
Simple dynamical considerations suggest that A2029 will accrete these subsystems in next few Gyr. 
In agreement with simulations and with other clusters observed in a similar redshift range, 
 the total mass in the A2029 infall region is comparable with the A2029 M$_{200}$ and will mostly be accreted in the long-term future. 
\end{abstract}
\section{INTRODUCTION}

Galaxy clusters grow hierarchically through accretion of generally lower mass systems.
More massive clusters typically form later than lower massive systems
 \citep{Neto07, BoylanKolchin09, McBride09}. 
Numerical simulations suggest that  
 the mass accretion rate is roughly proportional to the cluster mass:
 more massive clusters accrete more mass \citep{vandenBosch02, Fakhouri08, Fakhouri10, Giocoli12, deBoni16}. 

Observational estimates of the mass within the infall region of galaxy clusters 
 enable measurement of mass accretion rates \citep{Diaferio97, Rines02, deBoni16}. 
The measurement of the mass infall rate is challenging 
 because detailed observations covering the clusters outskirts (or infall regions) are required. 

Wide field-of-view redshift surveys, X-ray observations, 
 and weak lensing offer complementary views of the infall region of an individual cluster. 
So far, there are relatively few systems where all of these observations are available. 
Wide field-of-view redshift surveys \citep{Geller99, Reisenegger00, Rines02}
 apply the caustic technique \citep{Diaferio97, Diaferio99} to estimate the mass in the infall region. 
\citet{Rines13} show that the typical mass in the infall region is comparable with $M_{200}$ 
 ($= (4\pi/3)R_{200}^{3} 200 \rho_{crit}$).

Identification of X-ray emitting groups in the infall region provides 
 another probe of the future accretion by the cluster.  
The X-COP project \citep{Eckert17} surveys galaxy clusters with very deep {\it XMM} images to study infalling groups. 
\citet{Haines18} also conduct a systematic survey of X-ray groups in the infall region of 23 clusters at $z \sim 0.2$.   
They estimate that the galaxy clusters typically accrete $32\%$ of their mass by redshift 0 through the accretion of these surrounding X-ray groups. 

Weak gravitational lensing is another method for estimating the amount of mass in the outskirts of clusters 
 (e.g. \citealp{Geller13, Umetsu17}). 
Unlike the X-ray mass estimates, 
 weak lensing mass estimates are independent of the cluster dynamical state. 
The estimated mass in the infall region derived from weak lensing observations is consistent with 
 caustic estimates from dense redshift surveys \citep{Geller13}.

Combining these complementary probes strongly constrains the mass within the cluster and its infall regions. 
Each method of measuring the potentially infalling mass has limitations. 
For example, mass estimates based on the X-ray depend on the assumption of hydrostatic equilibrium.
Dense spectroscopy is critical. 
Without redshifts, association between extended X-ray emission and the main cluster is ambiguous.
Weak lensing mass estimates may be contaminated by 
 the presence of foreground/background systems (e.g. \citealp{Hoekstra11, Geller13, Hwang14}). 
The redshift survey facilitates the separation of cluster members from these foreground/background structures. 
  
Here, we combine spectroscopy, X-ray, and weak lensing observations
 to study the future mass accretion by the nearby massive cluster A2029. 
A2029 is one of the most massive clusters at $z = 0.079$. 
A2029 is well studied cluster with {\it ROSAT}, {\it XMM}, {\it Suzaku} and {\it Chandra} observations
 (e.g. \citealp{Lewis02, Clarke04, Walker12, PaternoMahler13}). 
\citet{McCleary18} construct weak lensing map of A2029. 
\citet{Sohn17a} conduct a redshift survey of this cluster (see also \citealp{Tyler13}). 
They examine the statistical properties of the A2029 member galaxies including luminosity, stellar mass, and velocity dispersion functions.

The complete redshift survey we discuss extends the survey of \citet{Sohn17a}. 
Based on complete spectroscopy,
 we investigate the core of A2029 and its infall region. 
We identify two relatively massive subsystems in the infall region and investigate their physical properties 
 based on spectroscopy, X-ray and weak lensing maps. 
In this process, we refine the X-ray estimates of the subsystem masses. 
We probe the future dynamical evolution of the A2029 system based on the physical properties of the infalling groups. 
  
We also use the complete redshift survey to make a census of foreground/background systems. 
Construction of this census is critical to removing ambiguous contributions to the mass within the infall region. 
Including A2029 and the two infalling groups, 
 we find a total of 13 extended X-ray sources. 
Their physical properties are consistent with the well-known scaling relation between X-ray luminosity and velocity dispersion.

The combined analysis we discuss sets the stage for future large datasets 
 including these complementary probes of the mass distribution in and around clusters of galaxies. 
eROSITA \citep{Merloni12}, Prime-Focus Spectrograph (PFS) on {\it Subaru} \citep{Takada14}, and Euclid \citep{Amendola18} 
 will provide these observations for clusters
 with a wide range of masses and redshifts thus tracing the detailed evolution of these systems.

We describe the redshift survey of A2029 in Section \ref{data}. 
We explain the identification of the cluster members using spectroscopic data in Section \ref{memsel}. 
In Section \ref{a2029struct}, 
 we identify two groups within the infall region of A2029 along with foreground/background systems in the A2029 field. 
We summarize their aggregate properties by placing them on the well-known $L_{X} - \sigma_{cl}$ scaling relation.
Finally, we discuss the past and future accretion history of A2029 (Section \ref{discussion})
 based on the dynamical connection between A2029 and the massive infalling groups. 
Throughout this paper, 
 we use the standard $\Lambda$CDM cosmology parameters: 
 $H_{0} = 70 ~\kmsmpc$, $\Omega_{m} = 0.3$ and $\Omega_{\Lambda} = 0.7$.

\section{REDSHIFT SURVEY OF A2029}\label{data}

Abell 2029 (R.A., Decl., z : 227.728729, 5.76716, 0.079) 
 is one of the most massive clusters in the local universe \citep{Sohn17a}.
A2029 was once known as a relaxed cluster 
 because of its smooth X-ray temperature profile (e.g. \citealp{Sarazin98}). 
However, recent deep X-ray observations reveal an X-ray sloshing spiral pattern 
 which indicates complex dynamical evolution \citep{Clarke04, PaternoMahler13}. 
Furthermore, the X-ray observations identify nearby extended X-ray sources 
 which may be galaxy systems that will eventually accrete onto the cluster \citep{Walker12}. 
To understand the dynamical status and the structure of A2029, 
 a dense redshift survey is important. 
Therefore, we extend the redshift survey for A2029 \citep{Sohn17a}. 
We include the total list of 4376 redshifts in Table \ref{allcat} of this Section.

\subsection{Photometry}

The galaxy catalog for the A2029 field is  
 based mainly on the Sloan Digital Sky Survey (SDSS) Data Release 12 (DR12, \citealp{Alam15}). 
Following \citet{Sohn17a}, 
 we first select extended objects with $r_{petro, 0} < 22$ mag
 within 100 arcmin of the cluster center. 
Extended objects have $probPSF = 0$, 
 where the $probPSF$ is the probability that the source is a star. 
We visually inspect extended sources 
 and remove some suspicious objects including stellar bleed trails and fragments of galaxies. 
In the north-eastern part of A2029, 
 there is a small patch where the SDSS DR12 photometry is missing. 
We supplement the A2029 galaxy catalog with the SDSS DR7 galaxy catalog for this region. 

Following \citet{Sohn17a}, 
 we use the $ugriz$ composite model (cModel) magnitudes, 
 a linear combination of de Vaucouleurs and model magnitudes. 
We apply the foreground extinction correction for each band. 
Throughout this paper, 
 all magnitudes indicate extinction corrected cModel magnitudes.
Our A2029 galaxy catalog contains 96082 extended objects 
 brighter than $r = 22$ mag within $R_{cl} < 100\arcmin$. 

\subsection{The Redshift Survey}

Based on the photometric galaxy catalog, 
 we conducted a spectroscopic survey of A2029. 
We first collect redshifts from previous redshift surveys including SDSS DR12. 
SDSS acquires spectra using $3\arcsec$ fibers for bright galaxies with $r < 17.77$. 
In the A2029 field, 
 3109 objects have SDSS redshifts with a typical uncertainty of $7~\kms$.
  
We also compiled 1308 redshifts from \citet{Tyler13} 
 who carried out a redshift survey using Hectospec 
 mounted on the MMT 6.5m telescope. 
Hectospec \citep{Fabricant05} is a 300 fiber-fed spectrograph, 
 which can obtain $\sim250$ spectra with a single exposure.  
\citet{Tyler13} obtained spectra of A2029 galaxies 
 and measure the redshifts and H$\alpha$ equivalent widths 
 to study star forming galaxy evolution in the cluster environment. 
The spectra taken from \citet{Tyler13} are available through the MMT archive
 \footnote{http://oirsa.cfa.harvard.edu/archive/search/}. 
From these spectra, 
 we measure the redshifts and visually inspect the redshift fits (see below for details)
 for consistency with the rest of our survey. 
Additionally, 
 we added 440 redshifts from the literature (e.g. \citealp{Bower88, Sohn15}) 
 through the NASA/IPAC Extragalactic Database (NED).
 
We conducted a deeper redshift survey of A2029 also using MMT/Hectospec. 
\citet{Sohn17a} report 982 redshifts of A2029 members. 
Here, we extend the redshift survey by including fainter objects. 
We use the 270 line mm$^{-1}$ Hectospec grating. 
The resulting spectra have 6.2 \AA~ spectral resolution and cover 3800 - 9100 \AA. 
The typical exposure time for each field is an hour. 

We used the IDL HSRED v2.0 package, developed by R.Cool and modified by MMT TDC, 
 to reduce the data. 
We measure the redshifts based on the cross-correlation of observed spectra 
 with a set of templates using RVSAO \citep{Kurtz98}. 
The cross-correlation results are visually inspected 
 and classified into three groups:
 `Q' for high-quality fits, `?' for ambiguous cases, and `X' for poor fits.  
We obtained a total of 2890 high-quality redshifts with a median redshift uncertainty of $32~\kms$; 
 1388 of these redshifts are new here compared to \citep{Sohn17a}. 
Among these, 97 objects are stars with $|cz| < 500~\kms$ (Appendix \ref{app}). 
 
There are 321 objects with both SDSS and MMT spectra.
The redshifts from SDSS and MMT for the duplicated objects have $|\Delta cz / (1 + z_{SDSS})| \leq 14~\kms$.  
For these objects, we use the redshift from the MMT.  

\begin{figure}
\centering
\includegraphics[scale=0.45]{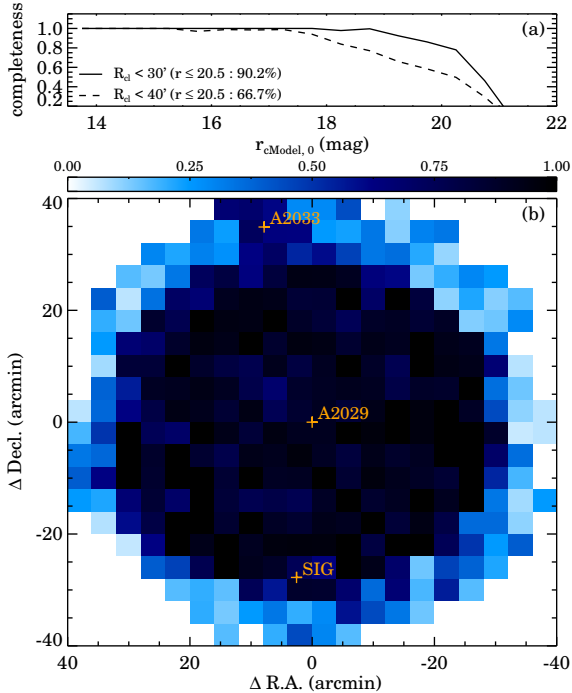}
\caption{
(a) Spectroscopic survey completeness as a function of $r-$band magnitude.
The solid and dashed lines plot 
 the completeness within $R_{cl} <~30\arcmin$ and $<~40\arcmin$, respectively. 
The survey is $\sim 90\%$ complete to $r_{petro, 0} = 20.5$ mag within $R_{cl} <~30\arcmin$. 
(b) Two-dimensional map of the redshift survey completeness to $r_{petro, 0} = 20.5$ mag 
 within $R_{cl} < 40\arcmin$. 
The yellow crosses mark the centers of A2033, A2029, and the southern substructure, SIG (from top to bottom). }
\label{zsurvey}
\end{figure} 

Figure \ref{zsurvey} displays the completeness of the A2029 redshift survey. 
Figure \ref{zsurvey} (a) shows the completeness as a function of $r-$band magnitude. 
We investigate the completeness within $R_{cl} < 30\arcmin$ and $<~40\arcmin$.
The integrated completeness to $r_{petro, 0} = 20.5$ mag within $R_{cl} < 30\arcmin$ is $90.0\%$
 (66.5\% within $R_{cl} < 40\arcmin$). 
Figure \ref{zsurvey} (b) shows the two-dimensional completeness map for the redshift survey. 
The redshift survey is uniformly $\gtrsim 90\%$ complete within $R_{cl} < 30\arcmin$. 
The survey completeness declines rapidly outside $R_{cl} = 30\arcmin$. 
Nevertheless, we use the redshift survey data within $40\arcmin$
 because Abell 2033 is included in this larger field of view. 
The yellow crosses in Figure \ref{zsurvey} (b) mark the positions of galaxy systems at the A2029 redshifts:
 A2029, A2033 and the southern infalling group (SIG, see below).
 
\subsection{The Redshift Catalog}

Table \ref{allcat} lists all of the spectroscopic redshifts within $40\arcmin$ of A2029
 including redshifts from our survey and from the literature. 
Table \ref{allcat} includes the SDSS object ID, Right Ascension, Declination, $r-$band cModel magnitude,
 redshift and its error, and the source of the redshift. 
The Table also includes the A2029 membership determined based on the caustics  
 (\citealp{Diaferio97}, see Section \ref{memsel}). 
In total, there are 4376 redshifts for galaxies in the field. 

\begin{deluxetable*}{lcccccccc}
\tablecolumns{9}
\tabletypesize{\scriptsize}
\tablewidth{0pt}
\setlength{\tabcolsep}{0.07in}
\tablecaption{The Redshift Catalog of A2029}
\tablehead{
\colhead{SDSS Object ID} & \colhead{R.A} & \colhead{Decl.} & \colhead{$r_{cModel, 0}$} & \colhead{z} & \colhead{z Source\tablenotemark{a}} & \colhead{Membership\tablenotemark{b}}}
\startdata
1237658780557836308 & 227.740259 &   5.766147 &  16.44 & $  0.07731 \pm   0.00010$ &    2,3 & Y \\
1237658780557836317 & 227.750323 &   5.756988 &  17.63 & $  0.08463 \pm   0.00009$ &      2 & N \\
1237658780557836338 & 227.737793 &   5.762320 &  19.07 & $  0.07593 \pm   0.00007$ &      2 & Y \\
1237658780557836305 & 227.744635 &   5.770809 &  16.45 & $  0.07455 \pm   0.00007$ &    2,3 & Y \\
1237658780557836311 & 227.738249 &   5.754465 &  17.90 & $  0.07726 \pm   0.00009$ &      2 & Y \\
1237658780557836336 & 227.732202 &   5.761856 &  18.47 & $  0.08085 \pm   0.00014$ &      1 & Y \\
1237658780557836342 & 227.749463 &   5.769346 &  18.83 & $  0.07828 \pm   0.00007$ &      2 & Y \\
1237658780557836316 & 227.735039 &   5.751555 &  17.46 & $  0.07921 \pm   0.00008$ &    1,2 & Y \\
1237658780557836337 & 227.732491 &   5.765348 &  19.37 & $  0.07899 \pm   0.00006$ &      2 & Y \\
1237658780557836369 & 227.731678 &   5.764879 &  20.01 & $  0.07735 \pm   0.00015$ &      1 & Y
\enddata
\label{allcat}
\tablecomments{
The entire table is available in machine-readable form in the online journal. 
Here, a portion is shown for guidance regarding its format. }
\tablenotetext{a}{The source of redshifts: (1) this survey, (2) \citet{Tyler13}, (3) SDSS, (4) NED, and (5) \citet{Sohn15}. }
\tablenotetext{b}{The membership determined based on the caustics. }
\end{deluxetable*}

\begin{figure}
\centering
\includegraphics[scale=0.45]{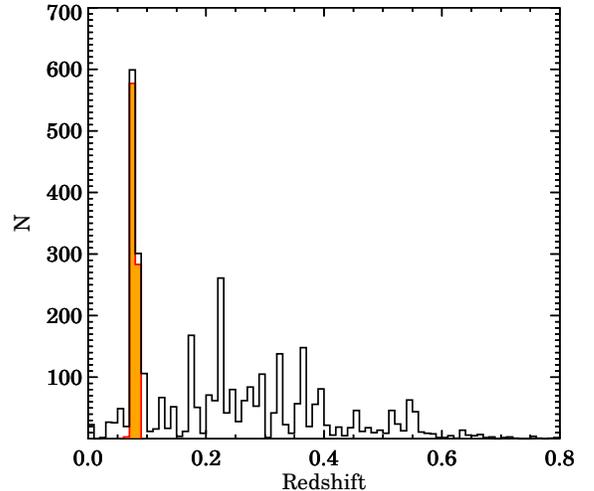}
\caption{
Redshift distribution of galaxies in the A2029 redshift survey. 
The filled histogram shows A2029 members identified based on the caustics. }
\label{zhist}
\end{figure}

Figure \ref{zhist} shows the redshift distribution of galaxies in the A2029 field.
The dominant peak in the distribution at $z \sim 0.08$ is A2029.  
Several less dominant peaks appear at higher redshift. 
These peaks include several readily identifiable background groups and one foreground group (see Section \ref{struct}). 

\begin{figure}
\centering
\includegraphics[scale=0.3]{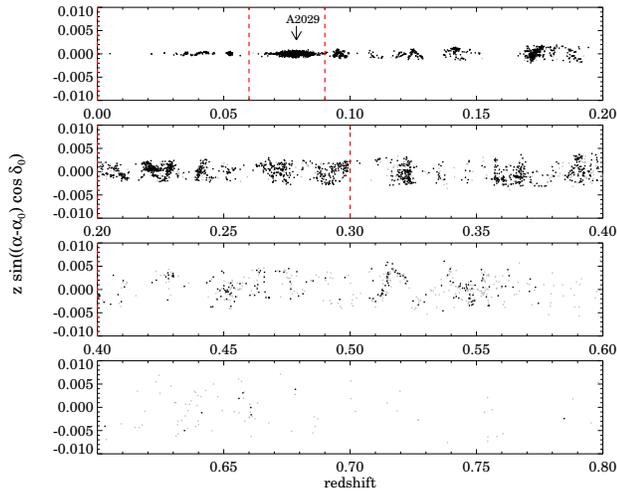}
\caption{
Cone diagram of the A2029 field within $R_{cl} < 40\arcmin$ projected in the R.A. direction.  
Black (gray) points are the galaxies brighter (fainter) than $r_{cModel, 0} = 20.5$. 
Dashed vertical lines display the boundaries we use for identifying 
 galaxy surface number densities that correspond to groups as described in Section \ref{struct}.}
\label{cone}
\end{figure}

Figure \ref{cone} displays the cone diagram for galaxies in the A2029 field 
 projected along the R.A. direction. 
Black points are the galaxies brighter than $r_{cModel, 0} = 20.5$ and 
 gray points are fainter galaxies. 
A2029 is the densest feature in the field. 
In the background of A2029, 
 the cone diagram shows the characteristic large-scale structure characterizes by 
 sizable voids and thin dense structures.  
Even corrected for the magnitude selection, 
 all of these background structures are much less dense that A2029. 

\section{MEMBER SELECTION}\label{memsel}

The caustic technique \citep{Diaferio97, Diaferio99, Serra13} is a powerful tool
 for identifying cluster members based on a spectroscopic survey.
The caustic technique calculates the escape velocity from the cluster
 and provides a mass profile of the cluster as a function of projected distance from the cluster center. 
As a by-product, 
 the technique identifies cluster members within the caustic pattern.  
\citet{Serra13} showed that 
 the technique successfully identifies $\sim95\%$ of cluster members within $3R_{200}$ 
 from mock catalogs with $\sim1000$ galaxies including $\sim 180$ members per cluster. 
The contamination from the interlopers is small: 
 $\sim 2\%$ within $R_{200}$ and $\sim8\%$ within $3R_{200}$. 

\begin{figure}
\centering
\includegraphics[scale=0.48]{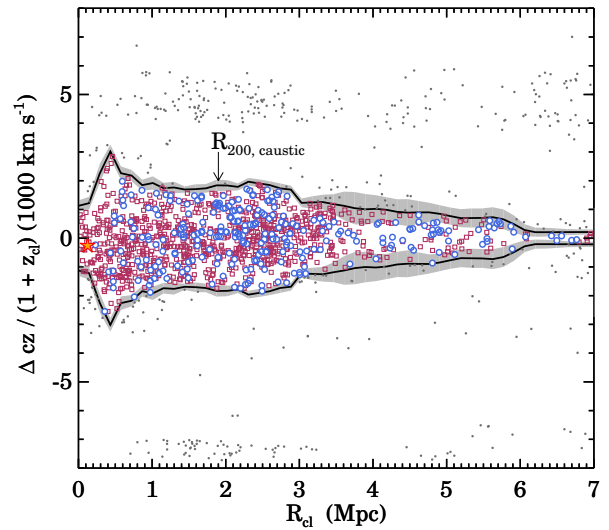}
\caption{R-v diagram for A2029. 
Gray dots are individual galaxies with redshifts. 
Red squares and blue circles are quiescent ($\dn > 1.5$) and star-forming ($\dn \leq 1.5$) members of A2029, respectively.
The star shows the BCG of A2029 (IC 1101). 
The rest frame line-of-sight velocity difference between the BCG and the cluster means is $-170 \pm 107~\kms$.
The arrow indicates $R_{200}$ derived from the caustic mass profile. }
\label{rv}
\end{figure} 

Figure \ref{rv} shows the relative rest-frame line-of-sight velocity difference versus 
 the projected distance for A2029, the R-v diagram.
In this phase space, a typical cluster shows a well-defined trumpet-like pattern \citep{Kaiser87, Regos89}. 
The solid lines show the caustics we derive based on the A2029 spectroscopic data
 and the shaded regions display the uncertainty in the caustic location. 
The caustics distinguish clearly between the cluster members and other galaxies along the line-of-sight. 
Gaps between cluster members and the foreground/background galaxies are intrinsic 
 to the large-scale galaxy distribution; 
 they do not originate from the incompleteness of the redshift survey. 

We derive the characteristic mass $M_{200}$ and radius $R_{200}$ based on the caustics:
 $M_{200} = 8.47^{+0.25}_{-0.23} \times 10^{14}$ M$_{\odot}$ and $R_{200} = 1.91^{+0.17}_{-0.19}$ Mpc. 
Here, $M_{200}$ and $R_{200}$ indicate the mass and the radius 
 where the mean density is 200 times the critical density of the universe. 
The derived $M_{200}$ and $R_{200}$ are consistent with the values we derived in \citet{Sohn17a} 
 based on a somewhat less complete redshift survey. 
The spectroscopically derived M$_{200}$ and $R_{200}$ are also compatible 
 with those based on the X-ray temperature profile \citep{Walker12}:
 the $M_{200, X-temp}$ is $8.0^{+0.15}_{-0.15} \times 10^{14}$ M$_{\odot}$ and 
 the $R_{200, X-temp}$ is $1.92^{+0.11}_{-0.13}$ Mpc. 
We also calculate the velocity dispersion of A2029 following the recipe given in \citet{Danese80}. 
We use the 571 members within $R_{cl} < R_{200}$. 
The velocity dispersion of A2029 is $967 \pm 25~\kms$, 
 consistent with the value derived in \citet{Sohn17a}.  

We identify 1215 spectroscopic members of A2029 within the caustics;
 571 cluster members are within $R_{200}$.   
The number of A2029 members exceeds  \citet{Sohn17a} who reported 982 spectroscopic members. 
A2029 is the one of the best sampled clusters.
There are 441 members in A2029 with $M_{r} < -18$ and $R_{cl} < R_{200}$; 
 Coma has 530 members with the same selection \citep{Sohn17a}. 
Hereafter, we refer to the 1215 spectroscopic members within the caustics as members of the A2029 system 
 and the 571 members within $R_{cl} < R_{200}$ as members of A2029. 

A2029 contains the most massive BCG (IC 1101) in the local universe \citep{Uson91}.
We investigate the projected distance between the BCG and the X-ray center following previous studies.
\citet{Patel06} reported that the BCG of A2029 is 131 kpc from the X-ray peak.
However, they used the X-ray center based on {\it ROSAT} data with a large PSF. 
\citet{Lauer14} demonstrated that 
 there are significant differences between the {\it ROSAT} X-ray centers and the {\it Chandra} X-ray centers 
 for a significant fraction of their cluster sample. 
Indeed, the offset between the A2029 BCG and the {\it Chandra} X-ray position is only 0.42 kpc 
 (cf. 1 kpc in \citealp{Lauer14}). 

Next, we examine the position of the A2029 BCG 
 with respect to the cluster center in redshift space (Figure \ref{rv}). 
The projected offset ($\Delta R_{cl}$) is $123 \pm 170$ kpc and 
 the radial velocity offset in the rest frame of the cluster is $-170 \pm 107~\kms$. 
The positional uncertainties include the error in the caustic center \citet{Serra11}.  
Within the uncertainty, IC 1101 lies on the kinematic center of A2029. 

\section{STRUCTURE OF THE MASSIVELY ACCRETING CLUSTER A2029}\label{a2029struct}

Current structure formation models suggest that 
 galaxy clusters grow hierarchically through the accretion of lower mass systems (e.g. \citealp{Bond96, Colberg05}).
Mass estimate for infalling groups 
 provide a direct estimate of the mass accretion rate (or the growth rate, \citealp{deBoni16, Haines18}). 
  
Systems around clusters can be identified independently with 
 X-ray, lensing, photometric, and spectroscopic observations.  
Systematic X-ray surveys show that 
 many local clusters have X-ray emitting groups in the infall regions \citep{Rines02, Haines18}. 
These X-ray emitting groups may be accreting systems (e.g. \citealp{Rines02, Haines18}).  
Gravitational lensing is another sensitive tool for identifying accreting groups
 by tracing the mass distributions in the cluster field \citep{Okabe10, Martinet16}. 
A dense spectroscopic survey enables the detection of lower mass systems around the cluster
 \citep{Yu15, Yu16, Yu18, Liu18}. 
This method has the advantage that the redshifts of the lower mass systems are known; 
 the X-ray and lensing candidate systems may not be at the main cluster redshift. 

Here, we search for galaxy groups around A2029 based on all of these methods (Section \ref{infalling}). 
Taking advantage of the redshift survey, 
 we construct a number density map of cluster members which facilitates group identification. 
We also utilize both the weak lensing map and X-ray observations of the cluster 
 to obtain physical properties of the groups.
We describe the identification of foreground/background groups in Section \ref{struct}. 

\subsection{Infalling Groups}\label{infalling}
\begin{figure*}
\centering
\includegraphics[scale=0.65]{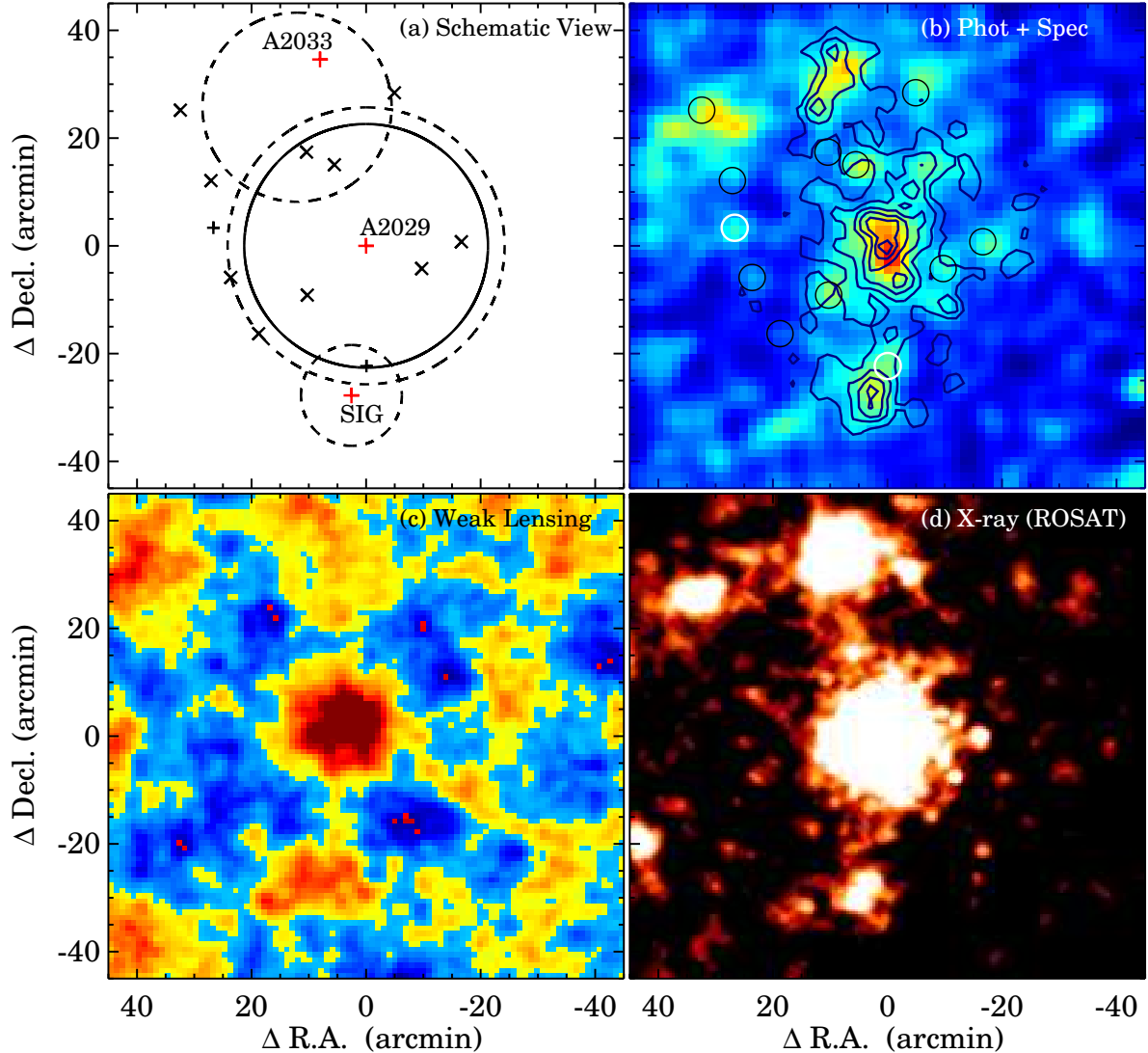}
\caption{Multi-wavelength view of A2029. 
(a) Schematic view of the A2029 field. 
Red pluses mark the position of A2029, Southern Infalling Group (SIG), 
 and the previously known center of A2033. 
Black crosses and pluses are the position of overdensities shown in the surface number density map of the galaxy distribution. 
Crosses have X-ray counterparts; pluses do not. 
The solid circle indicates $R_{200}$ derived from the caustic mass profile and 
 dashed circles are $R_{200}$ estimated from the $L_{X} - M_{200}$ relation from \citet{Leauthaud10}. 
(b) Galaxy distribution in the A2029 field (the colored map) compared with 
 the number density map of spectroscopically identified A2029 members (contours).
(c) Weak lensing significance map of A2029 based on DECam imaging \citep{McCleary18}.
 Red indicates high significance, blue indicates low significance. 
(d) ROSAT X-ray map of the A2029 region. }
\label{map}
\end{figure*}

Figure \ref{map}(a) is a schematic view of the A2029 field. 
A2029 is at the center of the field.
Red crosses mark the positions of possibly infalling groups associated with A2029. 
These groups correspond to peaks in the number density map of spectroscopic members. 
Abell 2033 is one of these groups; we display the A2033 center from NED. 
We also display the center of the Southern Infalling Group (SIG). 
Several additional galaxy overdensities 
 appear in the surface number density map of the spectroscopic survey; 
 black crosses/pluses show galaxy overdensities with/without X-ray counterparts  (see Section \ref{struct}).
The solid circle centered on A2029 is the $R_{200}$ based on the caustics. 
The dashed circles are the $R_{200}$ of A2029, A2033 and SIG measured based on their X-ray luminosities (Section \ref{prop}). 

Figure \ref{map}(b) shows a surface number density map for the spectroscopically identified members of A2029. 
For comparison, 
 we plot a background color map displaying
 the number density of photometric galaxies with $r \leq 20.5$ in the A2029 field. 
To avoid confusion, 
 we refer to the surface number density map of members and of photometric galaxies 
 as the member density map (contour) and the galaxy density map (color map), respectively. 

The member density map of A2029 is complex.
Overall, the distribution is elongated in the North-South direction.
At the northern edge, 
 a complicated structure includes A2033. 
The member density peak of A2033 is slightly offset from the peak in the photometric galaxy density map. 
The offset results from contamination by background galaxies (see Section \ref{struct}). 
The southern group matches the peak of the photometric galaxy density map. 
A2033 is seven times and SIG is four times more dense than the mean number density of spectroscopic members
 within a 500 kpc width annulus at similar clustercentric distance. 

Figure \ref{map}(c) shows a weak lensing map of the A2029 field. 
\citet{McCleary18} constructed a weak lensing map from $\sim 160,000$ galaxies 
 with shapes and 5-band photometric redshift estimates based on imaging with DECam.  
The full description of the weak lensing analysis is contained in \citet{McCleary18}; 
 here we provide only a brief summary.  

The A2029 field was observed in the {\it ugriz} filters in three runs between 2013 and 2015, 
 with total exposure times ranging from 7200s in {\it u} to 3200s in {\it r}.  
The shapes for weak lensing were derived from the {\it i} imaging where the mean seeing was 0.95\arcsec.   
The DECam data were processed using the NOAO Community pipeline\footnote{NOAO Data Handbook v2.2, Shaw et al. 2015} and 
 then stacks and PSF modeling was done using the THELI pipeline \citep{Erben01}. 
Galaxy shapes were computed using the KSB algorithm, 
 as implemented in \citet{Erben01} and \citet{vonderLinden14}.  
Galaxy photometry was calibrated by comparing photometric catalogs to the SDSS catalog of the sky, 
 and all galaxies with half-light radius greater than 1.15 and 
 the model PSF size brighter than the 50\% completeness limit ($i=24.4$ for A2029) are used 
 in the weak lensing analysis (a total of 160,256 galaxies).  
Photometric redshifts based on the galaxy colors are assigned using BPZ \citep{Benitez00},  
 and galaxies to which BPZ assigns a probability that $z>0.18$ (0.1 greater than the redshift of A2029) 
 greater than 80\% are used in the mapping.

To map out the projected mass distribution, we use the aperture mass statistic \citep{Schneider96} with
 a compensated filter \citep{Schirmer04} which provides an effective smoothing scale of 2.5\arcmin\ for the map.  
The significance of the signal at each pixel in the map is estimated by constructing $2\times 10^6$ randomized realizations 
 in which the shapes of the galaxies are randomly shuffled, 
 and then measuring how often the randomized signal exceeds the true signal (see \citealp{McCleary15}).  
These significance measures are converted to an equivalent $\sigma$ confidence for display in Figure \ref{map}(c).

The detection significance for a subclump is taken to be the highest significance pixel. 
For A2029, the signal is so strong that none of the randomized realizations showed 
 as strong a lensing signal indicating a detection at $>5\sigma$. 
For both A2033 and SIG, 
 the situation is complicated by the complex morphology in the lensing map, 
 where multiple pixels have similar significance.  
We choose to follow the "highest pixel" prescription and report a significance of $3.0\sigma$ for A2033 and $3.3\sigma$ for SIG.

The morphology of the weak lensing map is similar to the photometric galaxy density map. 
The strongest peak is the core of A2029.
A2033 and SIG are also detected with high significance. 
Other overdensities in the surface density map also correspond to low significance features in the lensing map. 
The consistency between the lensing map and the photometric galaxy density map is expected
 because the lensing map traces the cumulative projected mass density 
 along the line-of-sight and within the weak lensing kernel. 
\citet{Okabe10} show maps for other systems with similar qualitative correspondence.  

Figure \ref{map} (d) displays the ROSAT X-ray image of the A2029 field. 
We use the image from the ROSAT Position Sensitive Proportional Counter (PSPC) observation
 (program ID: rp800249, P.I.: C. Jones). 
The ROSAT PSPC data were reduced using the ROSAT Extended Source Analysis Software (ESAS, \citealp{Snowden94}). 
The image is  background subtracted and exposure corrected in the R47 band (0.44 - 2.04 keV, as shown in \citealp{Walker12}). 

Bright X-ray emission is present at the centers of A2029 and A2033.
These X-ray sources are listed in the ROSAT Brightest Cluster Sample \citep{Ebeling98}. 
There is also clear X-ray emission near the SIG (the second ROSAT PSPC catalog, \citealp{Rosat00}). 
\citet{Walker12} mentioned the existence of this southern extended X-ray source.

The spectroscopy, weak lensing, and X-ray images paint a consistent view of the structure 
 associated with A2029 within $R_{cl} < 40\arcmin$. 
The core of A2029 is a massive galaxy system associated with bright extended X-ray emission. 
Two groups, A2033 and SIG, appear in all three maps. 
Table \ref{sub} summarizes the positions of A2029 and these groups. 

A2033 was known as a separate cluster from A2029 \citep{Abell89}. 
Several cluster finding algorithms based on the SDSS photometric galaxy catalog also identify this cluster
 \citep{Wen09, Hao10, Szabo11, Wen12}. 
A2033 is also listed in the ROSAT X-ray Brightest Cluster Sample \citep{Ebeling98}.  
Based on redshifts from NED and SDSS spectroscopy, 
 \citet{Sifon15} identified $\sim 190$ spectroscopic members of A2033
 at a mean redshift of $z_{A2033} = 0.0796$. 
They computed $R_{200}~(= 1.89 \pm 0.14$ Mpc) and $\sigma_{200}~(= 911 \pm 69~\kms)$ based on these members. 

\begin{figure}
\centering
\includegraphics[scale=0.35]{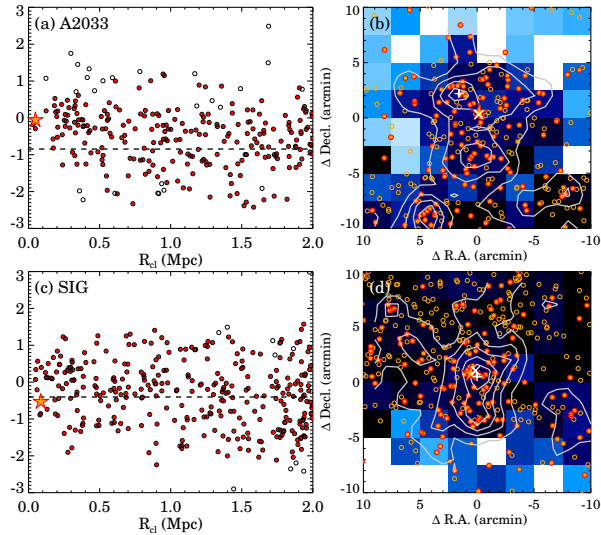}
\caption{
(a) R-v diagram centered on the A2033. 
 Open circles are the spectroscopic targets and 
  filled circles are the A2029 members within the caustics. 
 The star symbol indicates the brightest galaxy of A2033.
 The horizontal dashed line indicates the mean redshift of  A2029.  
(b) Galaxy distribution in the A2033 field. 
 The background map is the two-dimensional redshift completeness map. 
 Open circles and filled circles are spectroscopic targets and A2033 members, respectively. 
 The star symbol is the A2033 BGG. 
 The gray contours are the number density map of the A2033 members.
 The cross is the X-ray peak from the {\it Chandra} observations (Obs ID: 15167, P.I.: T.Reiprich)
 The plus symbol is the peak of the surface number density map of the A2033 members. 
(c)-(d) Same as (a) and (b), but for SIG. 
 The cross in panel (d) is the X-ray peak determined based on the {\it ROSAT} data. }
\label{ifg}
\end{figure} 

Figure \ref{ifg} (a) shows an R-v diagram centered on A2033. 
Within the $R_{200}$ from \citet{Sifon15} and $|\Delta cz / (1 + z_{A2033})| < 2000~\kms$, 
 there are 223 member galaxies. 
As expected from the member density map, 
 most A2033 members are within the A2029 caustics
 except for a few outliers with large velocity relative to the mean or A2029..  
The member distributions of A2029 and A2033 overlap in the phase-space diagram 
 making the identification of most members associated with A2033 ambiguous. 
The location of A2033 within the caustics suggests that it is dynamically connected to A2029 (see \citealp{Rines02}). 

Figure \ref{ifg} (b) displays the spatial distribution of the galaxies in the A2033 field. 
The background map shows the two-dimensional redshift survey completeness to $r_{cModel,0} = 20.5$. 
The open and the filled circles are 
 the spectroscopic targets and A2033 members, respectively.
The A2033 members have an elongated distribution (contour); 
 they are offset from both the brightest group galaxy (BGG) of A2033 (star symbol) and the {\it Chandra} X-ray center (cross). 
Because the redshift survey is homogeneous around the center of A2033, 
 the elongated distribution of the cluster members is not a product of incompleteness in the survey. 
Southeast of A2033,
 there is a loose concentration of galaxies at $z \sim 0.27$.  
Thus, the overdensity in the galaxy density map shown in Figure \ref{map} (b) 
 is contaminated by background galaxies. 
We conclude that the offset between the peak of the A2033 galaxy distribution and
 the A2033 BGG (or the X-ray center) is a physical property of the system. 
The mean redshift of A2033 members is $z_{A2033} = 0.0812$, 
 slightly larger than the mean redshift from \citet{Sifon15}. 
  
The BGG of A2033 is only $\sim5$ kpc from the {\it Chandra} X-ray center.  
The large offset (179 kpc) listed in \citet{Patel06} is 
 an overestimate due to the large uncertainty in the {\it ROSAT} X-ray center.  
In contrast to the X-ray, 
 the BGG of A2033 is certainly offset ($\sim 250$ kpc) from the peak surface number density of A2033 members 
 (panel (b) of Figure \ref{ifg}). 
The astrophysical implications of this offset are unclear. 
The radial velocity difference between the BGG and the A2033 mean is not significant ($\sim-57~\kms$).
  
We show the R-v diagram and the spatial distribution of the galaxies in the SIG field in Figure \ref{ifg} (c) and (d).   
There is no clear separation between the SIG members and the A2029 members. 
The brightest galaxy in SIG is significantly offset from the kinematic center of SIG. 
Furthermore, the spatial distribution of the SIG members is also elongated in the N-S direction 
as is A2033. The mean redshift of SIG is $z_{A2033} = 0.0802$. 
  
Measuring the offset between the BGG of SIG and the X-ray peak is challenging 
 because the {\it ROSAT} X-ray morphology of SIG  is disturbed. 
The northern peak in the {\it ROSAT} image corresponds to the BGG of SIG. 
The BGG offset from the northern X-ray peak is 38 kpc. 
The BGG is coincident with the surface number density peak of SIG members ($\Delta R_{cl} = \sim45$ kpc, panel (d) of Figure \ref{ifg}). 
Interestingly, the BGG is $\sim-533~\kms$ from the mean for SIG, a much larger difference than for the brightest galaxies of A2029 and A2033.  

We identify two galaxy overdensities associated with A2029 
 based on spectroscopy, weak lensing, and X-ray maps. 
A2033 and SIG appear in all three probes. 
We discuss the physical properties and implications of these systems for the future evolution of the A2029 system in Section \ref{accretion}. 
Additional overdensities not associated with A2029 appear in at least one of the maps; 
 we outline the properties of these systems in Section \ref{struct}.

\begin{deluxetable*}{lcccccc}
\tablecolumns{7}
\tabletypesize{\scriptsize}
\tablewidth{0pt}
\setlength{\tabcolsep}{0.05in}
\tablecaption{The Positions of A2029, A2033 and SIG}
\tablehead{
\colhead{ID} & \colhead{R.A.} & \colhead{Decl.} & \colhead{redshift} & 
               \colhead{BCG R.A.} & \colhead{BCG Decl.} & \colhead{BCG redshift}}
\startdata
A2029 & 227.728729 & +5.767164 & 0.0787 & 227.733751 & +5.744775 & 0.0778 \\
A2033 & 227.863556 & +6.340870 & 0.0812 & 227.860464 & +6.349078 & 0.0810 \\
SIG   & 227.771622 & +5.304444 & 0.0802 & 227.780800 & +5.317282 & 0.0783
\enddata
\label{sub}
\end{deluxetable*}

\subsection{Foreground/Background Groups in the A2029 Field}\label{struct}

The multi-wavelength maps reveal several foreground/background groups in the A2029 field. 
The cone diagram (Figure \ref{cone}), 
 the galaxy surface density map, 
 the weak lensing significance map, and the X-ray images in Figure \ref{map} 
 all show some concentrations of galaxies unassociated with the cluster. 

We examine the surface number density maps for galaxies in different redshift slices
 to identify foreground/background groups.  
The dashed vertical lines in Figure \ref{cone} indicate the boundaries of the six subsamples we consider. 
Each subsample includes a few probable groups. 

\begin{figure*}
\centering
\includegraphics[scale=0.60]{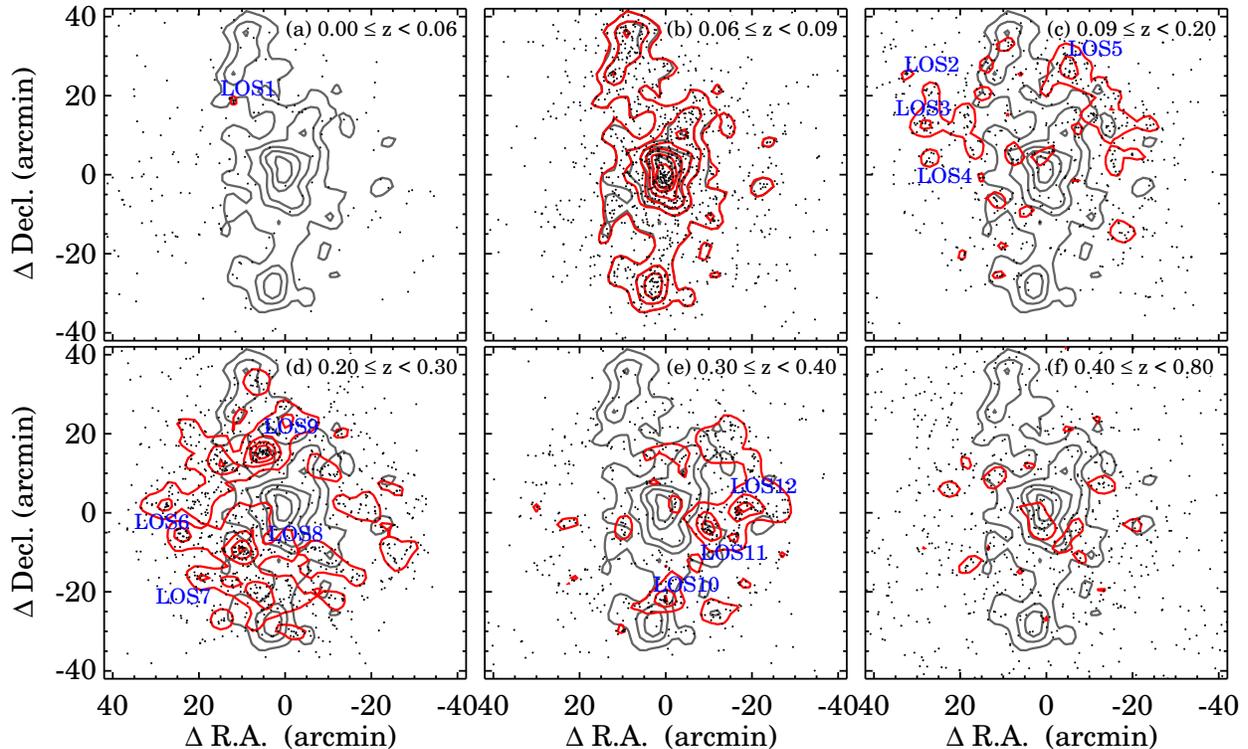}
\caption{
Red contours show  surface density maps for galaxies in different redshift ranges: 
(a) $0 \leq z < 0.06$, (b) $0.06 \leq z < 0.09$ (mainly A2029), (c) $0.09 \leq z < 0.20$, 
(d) $0.20 \leq z < 0.30$, (e) $0.30 \leq z < 0.40$ and (f) $0.40 \leq z < 0.80$.
Black points are individual galaxies. 
The gray contour shows the surface number density map for spectroscopic members of A2029. 
The lowest surface number density contours is 0.28 galaxies arcmin$^{-2}$; 
 the contours increase in steps of 0.28 galaxies arcmin$^{-2}$. }
\label{sdmap}
\end{figure*}

Figure \ref{sdmap} displays surface density maps for each of the redshift subsamples. 
Black points indicate individual galaxies in each subsamples and 
 red contours show the corresponding surface number density map. 
For comparison, 
 we show the surface number density map for spectroscopically identified cluster members (gray contours). 
The lowest level of the contours is 0.28 galaxies arcmin$^{-2}$ 
 and the contours increase in steps of 0.28 galaxies arcmin$^{-2}$. 

\begin{figure*}
\centering
\includegraphics[scale=0.75]{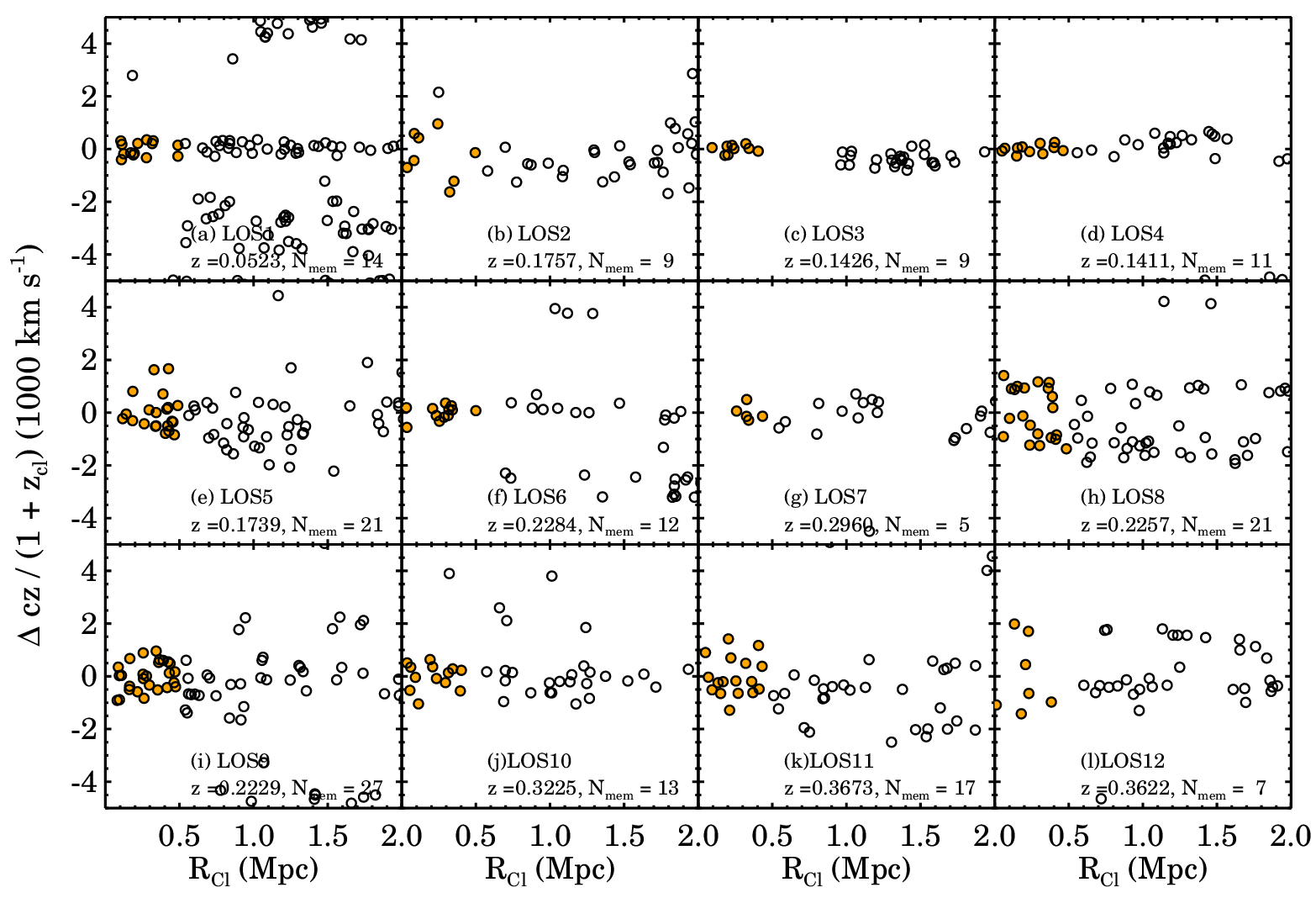}
\caption{R-v diagrams of the foreground/background groups in the A2029 field.
The open circles are spectroscopic targets. 
The orange filled circles are the members of the groups 
 within $R_{cl} < 5 \arcmin$ and $|\Delta cz / (1 + z_{cl}) < 2000~\kms$. }
\label{rvlos}
\end{figure*}

Figure \ref{sdmap}(a) shows the spatial distribution of objects with $0 \leq z < 0.06$. 
There are 157 galaxies in the foreground of the cluster.  
A small, tight group (LOS1, red contour) of 14 galaxies is at $z = 0.052$. 
The spectroscopic members are within the simple window: 
 $R_{cl} < 500$ kpc and $|\Delta cz / (1 + z_{cl})| \leq 2000~\kms$. 
We display the R-v diagram of this system in Figure \ref{rvlos}(a). 
The velocity dispersion of the system is $\sigma = 252 \pm 11~\kms$. 

We plot the galaxies with $0.06 \leq z < 0.09$ in Figure \ref{sdmap}(b). 
The galaxies in this redshift range are mostly cluster galaxies.
The surface number density map of the galaxies is essentially 
 identical to the cluster member density map. 
The contribution of non-cluster members 
 to the surface number density of the cluster member is negligible. 

Figure \ref{sdmap}(c) is based on 676 galaxies with $0.09 \leq z < 0.20$. 
We identify four groups (LOS2 - LOS5) with more than 9 members. 
These groups correspond to peaks in the photometric galaxy density map in Figure \ref{sub} (b), 
 but they are unrelated to A2029. 
The R-v diagrams of the groups are in panels (b) - (e) of Figure \ref{rvlos}. 
  
Figure \ref{sdmap}(d) shows the distribution of 1052 galaxies in the range $0.20 \leq z < 0.30$. 
We identify at least four groups (LOS6 - LOS9). 
Each group contains a significant number ($N > 12$) of members except LOS7. 
LOS7 is a superposition of galaxies at different redshifts. 
The other LOSs match galaxy overdensities in the photometric galaxy density map (Figure \ref{sub}(b)).  
 
LOS8 at $z = 0.226$ is a complicated structure including 21 spectroscopic members (Figure \ref{rvlos}(h)). 
The redshift distribution of LOS8 members is bimodal.
However, the spatial distributions of the galaxies in the two redshift groups are indistinguishable. 
LOS8 may be a group undergoing a merger.  
 
LOS9 with 25 members has associated X-ray emission.
Figure \ref{rvlos}(i) plots the R-v diagram for this group. 
\citet{Walker12} found X-ray emission near LOS9 based on ROSAT imaging data. 
They suggest that the X-ray source originates from 
 the overlap of X-ray emission from A2029 and A2033. 
However, the surface number density map indicates that 
 the X-ray emission is actually from a background system that appears as a finger in the cone diagram at $z \sim 0.223$.
The X-ray flux of this system from the ROSAT PSPC data is $6.17 \times 10^{-13}$ erg s$^{-1}$ cm$^{-2}$,  
 corresponding to an X-ray luminosity of $\sim 1.78 \times 10^{43}$ erg s$^{-1}$. 
The velocity dispersion of the system is $445 \pm 18~\kms$. 
The X-ray luminosity and the velocity dispersion of LOS9 are consistent with
 the $L_{X} - \sigma_{cl}$ relation derived from clusters at similar redshift \citep{Rines13}.  
 
Our redshift survey includes 688 galaxies with $0.30 \leq z < 0.40$. 
The projected spatial distribution of these galaxies is shown in Figure \ref{sdmap}(e). 
We identify three groups within this subsample.
LOS10 at $z = 0.326$ lies between A2029 and SIG. 
This group may impact mass estimate of A2029 based on weak lensing.
LOS11 with 17 members is located east of the cluster. 
LOS12 is a superposition of galaxies containing a group of six galaxies at $z = 0.362$. 
  
We plot the location of the 596 galaxies in the wide redshift range $0.40 \leq z < 0.80$
 in Figure \ref{sdmap}(f).  
The redshift survey appears sparse 
 because it includes only the most intrinsically luminous galaxies at this redshift. 
Thus we cannot identify background groups in this range. 

Table \ref{loscat} summarizes the foreground/background groups we identify in the A2029 field. 
Table \ref{loscat} includes the central position, redshift, 
 the number of spectroscopic members, the rest frame line-of-sight velocity dispersion, and 
 the X-ray luminosities. 
Here, the X-ray luminosities are measured within the energy band 0.1 - 2.4 keV based on the ROSAT image. 
The identification of these structures in the cluster field enables
 further understanding of the multi-wavelength view of the cluster. 
For example, the apparent X-ray emission between A2029 and A2033 is most probably flux from a background system at redshift $z = 0.223$. 

\begin{deluxetable*}{lcccccc}
\tablecolumns{6}
\tabletypesize{\footnotesize}
\tablewidth{0pt}
\setlength{\tabcolsep}{0.02in}
\tablecaption{Structures in the A2029 Field} 
\tablehead{
\colhead{ID} & \colhead{R.A.}  & \colhead{Decl.} & \colhead{redshift} & 
\colhead{$N_{mem}$\tablenotemark{a}} & \colhead{$\sigma$} & \colhead{$L_{X}$} \\
\colhead{}   & \colhead{(deg)} & \colhead{(deg)} & \colhead{}         & 
\colhead{}                           & \colhead{($\kms$)} & \colhead{($10^{43}$ erg s$^{-1}$)} }
\startdata
 LOS1 & 227.902850 &   6.057051 & 0.052 & 14 & $ 252 \pm  11$ & $ 0.0453 \pm  0.0041$ \\
 LOS2 & 228.271960 &   6.186980 & 0.176 &  9 & $ 993 \pm  41$ & $13.1195 \pm  0.3400$ \\
 LOS3 & 228.182250 &   5.968993 & 0.143 &  9 & $ 130 \pm   6$ & $< 0.0111$ \\
 LOS4 & 228.175492 &   5.823115 & 0.141 & 11 & $ 133 \pm   7$ & $< 0.0125$ \\
 LOS5 & 227.645900 &   6.240714 & 0.174 & 21 & $ 597 \pm  16$ & $ 3.3707 \pm  0.1606$ \\
 LOS6 & 228.124521 &   5.669391 & 0.228 & 12 & $ 212 \pm  10$ & $< 0.1072$ \\
 LOS7 & 228.042953 &   5.495509 & 0.296 &  5 & $ 230 \pm   9$ & $< 0.0146$ \\
 LOS8 & 227.900588 &   5.614726 & 0.226 & 21 & $ 789 \pm  25$ & $ 3.8967 \pm  0.2075$ \\
 LOS9 & 227.821348 &   6.018032 & 0.223 & 27 & $ 445 \pm  18$ & $ 3.1008 \pm  0.2119$ \\
LOS10 & 227.727834 &   5.395346 & 0.326 & 13 & $ 364 \pm  12$ & $ 0.8018 \pm  0.0773$ \\
LOS11 & 227.565872 &   5.696886 & 0.367 & 17 & $ 535 \pm  19$ & $ 7.0670 \pm  0.6234$ \\
LOS12 & 227.449909 &   5.779819 & 0.362 &  7 & $1021 \pm  35$ & $17.2136 \pm  0.8321$
\enddata
\label{loscat}
\tablenotetext{a}{Number of spectroscopic members within $R_{cl} < 500$ kpc and $|\Delta cz / (1 + z_{cl})| \leq 1000~\kms$
 from the center of the structures. }
\end{deluxetable*}

\section{The Past and Future of A2029}\label{discussion}

The dynamical history of A2029 is complex.
Deep {\it Chandra} X-ray observations reveal  
 an astonishing sloshing pattern extending to 400 kpc around the BCG \citep{Clarke04, PaternoMahler13}. 
Hydrodynamic simulations show that 
 the sloshing pattern can form through the interaction 
 between the cluster and an infalling subcluster \citep{ZuHone10}. 
By comparing the observed sloshing pattern with hydrodynamic simulations, 
 \citet{PaternoMahler13} suggest that
 A2029 interacted with an infalling group with 20\% of mass of A2029 ($\sim 1.7 \times 10^{14}$ M$_{\odot}$) between 2 and 3 Gyr ago. 

Based on the maps in Figure \ref{sub}, 
 we identify two galaxy groups at the same redshift as A2029. 
The members of these groups are within the caustic profile of A2029
 indicating that they will produce additional accretion events over the long-term future of the system. 

\citet{Haines18} identify similar infalling X-ray groups within the caustics of the primary clusters 
 in the LoCuSS cluster sample. 
\citet{Rines02} also identify several X-ray groups within the caustics of A2199. 
They compute the turnaround radius of A2199 ($\sim 6.4 - 8.1$ Mpc) and 
 identify three X-ray groups within the turnaround radius and the caustic as infalling groups. 
The two galaxy groups of A2029 we identify are well within the turnaround radius of A2029 
 ($\sim 10$ Mpc, larger than for A2199 because of its larger mass). 

In Section \ref{prop}, 
 we discuss the properties of A2033 and SIG. 
In Section \ref{twomodel}, we consider their probable future accretion by A2029. 
Section \ref{accretion} discusses the accretion in 
 a broader context including comparison of the group masses 
 with the total amount of material in the infall region.  

\subsection{The Physical Properties of the Infalling Groups}\label{prop}

We estimate the physical properties of the infalling groups including their membership, size, and mass. 
We determine the membership based on spectroscopy. 
Because of the proximity of the infalling groups to the cluster core,
 we cannot compute caustics for the groups free of contamination by members of the primary cluster. 
Thus, we identify group members by applying simple cuts:
 $R_{proj} < 500$ kpc and $|\Delta cz / (1 + z_{cl})| < 2000~\kms$. 
The projected radius cut is small enough 
 not to overlap $R_{200}$ of the primary cluster. 
The line-of-sight relative velocity criterion is comparable with the amplitude of the A2029 caustic 
 at the group position.
There are 57 and 70 spectroscopically identified members in A2033 and SIG, respectively. 
The typical numbers of projected A2029 members around A2033 and SIG distances are 8 and 18, respectively. 

The velocity dispersions of A2033 and SIG are 
 $701 \pm 74 ~\kms$ and $745 \pm 62~\kms$, respectively.  
To evaluate the velocity dispersion and its error for A2023 and SIG, 
 we use 1000 randomly selected subsets of 49 A2033 members and 52 SIG members.  
This process accounts for the average contamination by A2029 in the annulus. 
The velocity dispersion error indicates the $1\sigma$-deviation from 1000 velocity dispersion estimates. 
We derive $M_{200, \sigma}$s and $R_{200, \sigma}$s of the groups 
 based on the $M_{200} - \sigma$ scaling relation from \citet{Rines13}:
\begin{equation}
M_{200, \sigma} [10^{14} M_{\odot}] = 0.093 \times (\sigma / [200~\kms])^{(2.90 \pm 0.15)}.
\end{equation}
The estimated mass is $(3.50 \pm 2.21) \times 10^{14}$ M$_{\odot}$ for A2033 and 
 $(4.32 \pm 2.38) \times 10^{14}$ M$_{\odot}$ for SIG. 
We list the estimated $M_{200, \sigma}$ and $R_{200, \sigma}$ in Table \ref{subprop}. 

Next, we measure the masses of the systems based on the weak lensing profile. 
The weak lensing mass estimates follow the procedure developed in \citet{McCleary18}
 to simultaneously fit multiple mass components taking advantage of the linearity of the lensing deflections (as opposed to the shears).  
Each component is modeled as an NFW profile centered on the coordinate defined by the X-ray peak.  
For each component, the value of the gravitational lensing deflection as a function of the component $M_{200}$
 (the NFW concentration was fixed at 4 given the strong mass-concentration degeneracy) is then 
 derived at the position of each background galaxy using the photometric redshift of the galaxy and
 the cluster to define the distance ratios.
We numerically compute the gradient of the deflection to calculate the predicted shear at the position of each galaxy, 
 then vary the masses of the components to minimize the RMS difference 
 between the observed galaxy ellipticity tensors and the predicted shears.   
We estimate the uncertainty in the mass determinations 
 by measuring the scatter in the best-fit estimates made 
 by randomly selecting half the background galaxies as targets and repeating the fitting procedure for 500 sampled realizations.

In the weak lensing map, 
 A2029 is obviously the most massive system ($M = (9.6 \pm 1.8) \times 10^{14}$ M$_{\odot}$). 
The weak lensing mass estimate for A2033 is $(2.4 \pm 1.6) \times 10^{14}$ M$_{\odot}$ and
 the mass estimate for SIG is $(1.3 \pm 1.5) \times 10^{14}$ M$_{\odot}$. 
These mass estimates are smaller than the masses inferred from the $M_{200} - \sigma_{cl}$ relation. 
Interestingly, 
 the estimated mass based on the weak lensing profile of SIG is smaller than the mass of A2033. 

We obtain the X-ray properties from the {\it ROSAT} and {\it Chandra} data. 
Average temperature measurements were obtained 
 using X-ray spectral fitting of the {\it Chandra} data for Abell 2029 (obsids:892,4977,6101) and Abell 2033 (obsid:15167). 
The {\it Chandra} data were reduced using the latest version of CIAO (version 4.10) with spectra extracted 
 using the tool DMEXTRACT and response files created using MKWARF and MKACISRMF. 
X-ray fitting was performed using XSPEC with an absorbed APEC model \citep{Smith01}.
To obtain the X-ray luminosities of the A2029, A2033, SIG and LOS1-12, 
 we used the growth curve analysis method described in \citet{Bohringer13}.

A2029 is very bright in the X-ray with a luminosity of $(94.85 \pm 0.47) \times 10^{43}$ erg s$^{-1}$
 with a X-ray temperature of $\sim 7.5 keV$ \citep{Ebeling98, Walker12}. 
A2033 is much fainter than A2029; 
 the X-ray luminosity is $(17.69 \pm 0.19) \times 10^{43}$ erg s$^{-1}$ and the temperature is 3.7 keV. 
The X-ray temperature we derive for A2033 is $\sim1$ keV lower than the previous measurement quoted in the BCS catalog \citep{Ebeling98}. 

The X-ray morphology of SIG is disturbed and there are some X-ray point sources which contaminate 
 the X-ray emission. 
We measure the X-ray luminosity after subtracting the X-ray point sources: 
 the X-ray luminosity of SIG is then $(1.57 \pm 0.05) \times 10^{43}$ erg s$^{-1}$. 
This X-ray luminosity is only $\sim 40\%$ of the luminosity estimated from the flux 
 listed in the second ROSAT PSPC source catalog \citep{Rosat00}.
We are not able to estimate the temperature of SIG due to its low flux. 

\begin{figure}
\centering
\includegraphics[scale=0.48]{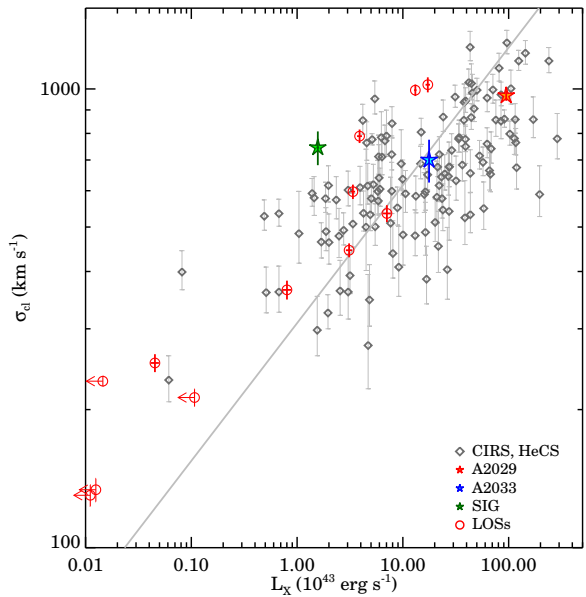}
\caption{Velocity dispersion ($\sigma_{cl}$) vs. X-ray luminosity ($L_{X}$) for A2029 (red star), A2033 (blue star) and SIG (green star). 
The red circles display the foreground and background systems in the A2029 field. 
The red circles with arrows are the systems where we measure only the X-ray upper limits. 
The gray diamonds show the CIRS \citep{Rines06} and HeCS \citep{Rines13} cluster samples. 
The solid line is the best-fit relation for nearby X-ray cluster samples \citet{Zhang11}.}
\label{lxsig}
\end{figure} 

Figure \ref{lxsig} displays 
 the velocity dispersion versus rest-frame X-ray luminosity within $0.1-2.4$ keV for A2029, A2033, and SIG. 
We also include the foreground and background systems in the A2029 field. 
For comparison, we add cluster samples from \citet{Rines06} and \citet{Rines13}. 
The solid line in Figure \ref{lxsig} shows
 the scaling relation for local clusters from \citet{Zhang11}.
 
A2029 and A2033 follow the $L_{X} - \sigma_{cl}$ relation 
 defined by previous cluster samples. 
SIG has a large velocity dispersion compared to the systems with similar X-ray luminosities, 
 but it still follows the $L_{X} - \sigma_{cl}$ relation within the velocity dispersion uncertainty. 
The scaling relation for all of the systems in the A2029 field is consistent with the local scaling relation. 
This consistency is a strong check of the combined analysis we present. 

We derive $M_{200, X}$ from the X-ray temperatures and luminosities of A2029, A2033, and SIG. 
We first compute the $M_{200, X{\rm temp}}$ of A2029 and A2033 
 based on the scaling relation of \citet{Arnaud05}:
\begin{equation}
M_{200, X-temp} = (5.74 \pm 0.3) \times 10^{14} (\frac{kT}{5 keV})^{(1.49 \pm 0.17)} / E(z)^{1/2},
\end{equation} 
where $E(z) = [\Omega_{m} (1 + z)^{3} + (1 - \Omega_{m})]$. 
The $M_{200, X-temp}$ of A2029 is $(10.25 \pm 0.85) \times 10^{14}$ M$_{\odot}$ \citep{Walker12} and 
 the $M_{200, X-temp}$ of A2033 is $(3.58 \pm 0.11) \times 10^{14}$ M$_{\odot}$. 
The mass estimates of A2029 and A2033 based on X-ray temperatures are close to 
 their mass estimates based on both velocity dispersion and weak lensing.

Next, we derive $M_{200, X-lum}$ using the scaling relation between $M_{200}$ and $L_{X}$ from \citet{Leauthaud10}. 
The $M_{200, X}$ of A2029 is $(1.75 \pm 0.08) \times 10^{15} M_{\odot}$, 
 a factor of two larger than the more robust mass estimates 
 based on caustic and the X-ray temperature profiles \citep{Walker12}. 
Thus, the masses derived for A2033 and SIG from the $M_{200} - L_{X}$ relation, even though they are widely used, must be regarded with some caution. 
The $M_{200}$s of A2033 and SIG are $(5.14 \pm 0.59) \times 10^{14} M_{\odot}$ and 
 $(0.32 \pm 0.02) \times 10^{14} M_{\odot}$, respectively. 
We convert $M_{200}$s into $R_{200}$s using the relation: 
 $M_{200} = 200 \rho_{crit}(z) (4\pi / 3) R_{200}^3$. 
The dashed circles in Figure \ref{map}(a) display the derived $R_{200}$s. 

Table \ref{subprop} summarizes all of the mass estimates for A2029, A2033, and SIG 
 based on the various proxies.  
For A2029, only the mass estimate based on the X-ray luminosity disagrees with other mass estimates. 
The mass estimates of A2033 from velocity dispersion, weak lensing, X-ray temperature agree within $1\sigma$. 
Again, the $M_{200, X-lum}$ of A2033 is larger than the other mass estimates. 
For SIG, the mass estimates based on weak lensing and X-ray luminosity are comparable. 
The larger offset of BGG with respect to the group mean redshift suggest that 
 the large $M_{200, \sigma}$ of SIG is unreliable.
In the following, 
 we use the mean of mass estimates based on velocity dispersion, weak lensing, and X-ray temperature (if available). 
We do not include the mass estimate based on the X-ray luminosity because it deviates substantially
 for the best determined cases, A2029 and A2033. 
 
\begin{deluxetable*}{lccccccccc}
\tablecolumns{10}
\tabletypesize{\scriptsize}
\tablewidth{0pt}
\setlength{\tabcolsep}{0.05in}
\tablecaption{The Physical Properties of A2029, A2033, and SIG}
\tablehead{
\colhead{ID}            & \colhead{N$_{mem}$\tablenotemark{a}} & 
\colhead{$\sigma_{cl}$} & \colhead{$M_{200, \sigma}$\tablenotemark{b}} & \colhead{$M_{WL}$} & 
\colhead{kT} & \colhead{$M_{200, X-temp}$\tablenotemark{c}} & 
\colhead{$L_{X}$} & \colhead{$M_{200, X-lum}$\tablenotemark{d}} & \colhead{$M_{200, mean}$\tablenotemark{e}} \\
\colhead{}   & \colhead{}                           & 
\colhead{($\kms$)}      & \colhead{($10^{14} M_{\odot}$)}               & \colhead{($10^{14} M_{\odot}$)} & 
\colhead{(keV)}    & \colhead{($10^{14} M_{\odot}$)}  & 
\colhead{($10^{43}$ erg s$^{-1}$)} & \colhead{($10^{14} M_{\odot}$)}   &
\colhead{($10^{14} M_{\odot}$)}  }
\startdata
A2029 & 597 & $967 \pm 25$ & $9.0 \pm 3.4$ & $9.6 \pm 1.8$ & 7.5      & $10.87 \pm 0.93$ & $94.85 \pm 0.47$ & $17.53 \pm  0.06$ & $9.8 \pm 1.3$ \\
A2033 &  57 & $701 \pm 74$ & $3.5 \pm 2.3$ & $2.4 \pm 1.6$ & 3.7      & $ 3.58 \pm 0.11$ & $17.69 \pm 0.19$ & $ 5.79 \pm  0.04$ & $3.2 \pm 0.9$ \\
SIG   &  70 & $745 \pm 62$ & $4.2 \pm 2.3$ & $1.3 \pm 1.5$ &  \nodata & \nodata          & $ 0.22 \pm 0.02$ & $ 1.17 \pm  0.02$ & $2.8 \pm 1.4$
\enddata
\label{subprop}
\tablenotetext{a}{The number of spectroscopic members within $R_{200}$ for A2029, and within 500 kpc for A2033 and SIG. }
\tablenotetext{b}{$M_{200, \sigma}$ estimated based on the $M_{200} - \sigma$ relation in \citet{Rines13}. }
\tablenotetext{c}{$M_{200, X-temp}$ estimated based on the $M_{200} - T_{X}$ relation in \citet{Arnaud05}. }
\tablenotetext{d}{$M_{200, X-lum}$ estimated based on the $M_{200} - L_{X}$ relation in \citet{Leauthaud10}.}
\tablenotetext{e}{The mean $M_{200}$ estimated from $M_{200, \sigma}$, $M_{200, WL}$, and $M_{200, X-temp}$ (if available).  }
\end{deluxetable*}

\subsection{Accretion of the Infalling Groups}\label{twomodel}

\begin{figure}
\centering
\includegraphics[scale=0.39]{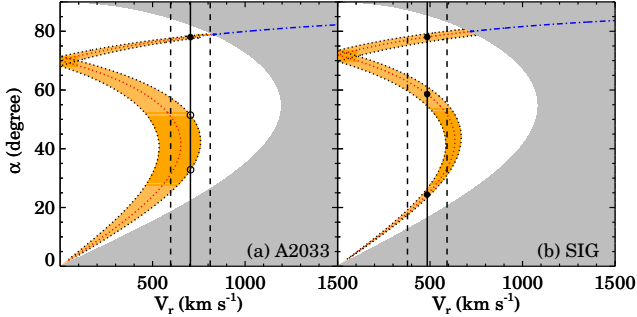}
\caption{
(a) Projection angle ($\alpha$) vs. radial velocity difference ($V_{r}$) of A2033
 given by the simple two-body model with M$_{tot} = 8.47 \times 10^{14}$ M$_{\odot}$ of A2029 \citep{Sohn17a}. 
The vertical solid and dashed lines show 
 the radial velocity differences and their 1$\sigma$ uncertainties. 
The open and shaded region indicate gravitationally bound and unbound region
 derived by the Newtonian criterion.
The dotted and dot-dashed lines plot the bound and unbound solutions, respectively. 
The orange shaded regions indicate the uncertainty in the bound solutions 
 originating from the mass estimate uncertainties. 
Filled circles indicate the solutions summarized in Table \ref{two}.
(b) Same as (a), but for SIG. }
\label{twobody}
\end{figure} 

To estimate the future accretion time of A2033 and SIG by A2029, 
 we apply a two-body model separately for the orbits of A2033 and SIG relative to A2029 \citep{Beers82}. 
This model computes a linear orbit for the system assuming there is no shear or net rotation. 
Following \citet{Beers82}, 
 we assume that A2033, SIG and A2029 are at zero separation at $t=0$
 and they are now moving away or approaching each other for the first time in their history. 
The equation of motion for this system is:
\begin{equation}
R = \frac{R_{p}}{\cos \alpha} = \frac{R_{m}}{2}(1 - \cos \chi), 
\end{equation}
\begin{equation}
V = \frac{V_{r}}{\sin \alpha} = (\frac{2GM}{R_{m}})^{1/2} \frac{\sin \chi}{(1 - \cos \chi)},
\end{equation}
\begin{equation}
t =(\frac{R_{m}^{3}}{8GM})^{1/2} (\chi - \sin \chi),
\end{equation}
 where $R_{p}$ is the projected distance from the main cluster and the substructure,  
 $\alpha$ is a projection angle between the plane of the sky and the line connecting two systems,
 $R_{m}$ is the separation of the systems at maximum expansion, 
 $\chi$ is the development angle, 
 $V_{r}$ is the relative radial velocity difference of the two systems, 
 and $M$ is the total mass of the system. 
 
To solve the equation of motion, 
 we use the observed $R_{p}$ and $V_{r}$ for the two groups: 
 $R_{p} = 3.17$ Mpc and $V_{r} = 704.5~\kms$ for A2033 and
 $R_{p} = 2.49$ Mpc and $V_{r} = 484.9~\kms$ for SIG. 
We use the dynamical mass of A2029 measured from the caustic and the X-ray temperature profile, 
 i.e. $M_{200} = 8.47 \times 10^{14} M_{\odot}$, 
 and the masses of A2033 and SIG derived from their X-ray luminosities. 
When we solve the equation, we use the sum of the A2029 mass and the mass of either A2033 or SIG. 
We also set $t = 12.8$ Gyr, the age of the universe at the redshift of cluster.
We solve the equation of motion by increasing $\chi$ from 0 to $2 \pi$. 
 
Figure \ref{twobody} shows the projection angle ($\alpha$) 
 as a function of radial velocity difference ($V_{r}$)
 of A2033 and SIG.
We first plot the Newtonian criterion for gravitational binding \citep{Beers82}:
\begin{equation}
V_{r}^{2} R_{p} \leq 2 G M_{\rm tot}~\sin^{2} \alpha~\cos \alpha. 
\end{equation}
This criterion divides gravitationally bound (open) and unbound (shaded) regions in Figure \ref{twobody}. 

By solving the equation of motion,
 we obtain a bound-outgoing solution for A2033.
According to this bound-outgoing solution, 
 A2033 is now at $\sim 13.5$ Mpc from A2029 and moves away with a velocity of $V \sim 725~\kms$.
This solution coincides with the result from \citet{Gonzalez18}. 
Considering the uncertainty in mass estimates of the groups and the mean redshift measurements, 
 there are also bound-incoming solutions (open circles in Figure \ref{twobody}(a)). 
For example, if we take the maximum total mass of this system, 
 A2033 is approaching A2029 at $V \sim 842~\kms$ or at $V \sim 1214~\kms$. 
In these case, A2033 collides with A2029 within $\sim 5.52$ Gyr or $\sim 2.84$ Gyr, respectively.
  
We also derive two bound-incoming and one bound-outgoing solutions for the A2029-SIG system. 
The bound-outgoing solution proposes that 
 SIG is moving away at $V \sim 497~\kms$ at a distance of $\sim 11.5$ Mpc.  
The bound-incoming solutions suggest that 
 SIG is approaching A2029 at $V \sim 581~\kms$ or at $V \sim 1111~\kms$. 
The estimated collisional time scale for each case is $\sim 7.59$ Gyr or $\sim 2.43$ Gyr, respectively. 
The relative probabilities ($p_{i}$) of these solutions is: 
\begin{equation}
p_{i} = \int^{\alpha_{sup,i}}_{\alpha{inf,i}} {\rm cos}~\alpha~{\rm d}\alpha, 
\end{equation}
 where $\alpha_{sup,i}$ and $\alpha_{inf, i}$ are computed 
 by taking into account the uncertainties in the total mass of the system and the radial velocity difference. 
Then, the relative probabilities are normalized by $\sum (p_{i})$. 
In the case of SIG, the probability that it is now incoming is $\sim 94\%$.   
Table \ref{two} summarizes these solutions. 

The detailed dynamics of the multi-component A2029 system is probably much more complicated.
The two-body model provides a guide to the timescales in the problem that complements 
 the fact that the groups lie within the caustics. 
The probability is roughly $\sim 70\%$ of a merger in the next 3 Gyr (e.g. one of the bound-incoming solution for SIG). 

\begin{deluxetable*}{lcccccccccc}
\tablecolumns{11}
\tabletypesize{\footnotesize}
\tablewidth{0pt}
\setlength{\tabcolsep}{0.05in}
\tablecaption{Two-body Model Solutions}
\tablehead{
\colhead{Substructure} & \colhead{$R_{p}$}   & \colhead{$V_{r}$}   & \colhead{Solution}  &
\colhead{$\chi$}       & \colhead{$\alpha$}  & 
\colhead{$R$}          & \colhead{$R_{max}$} & \colhead{$V$}       & \colhead{$t_{scl}$} & \colhead{P} \\
\colhead{}             & \colhead{(Mpc)}     & \colhead{($\kms$)}  & \colhead{}          &
\colhead{(rad)}        & \colhead{(deg)}     & 
\colhead{(Mpc)}        & \colhead{(Mpc)}     & \colhead{($\kms$)}  & \colhead{(Gyr)}     & \colhead{(\%)}}
\startdata
A2033 & 3.17 & 704.5 & Bound-Outgoing                  & 1.22 & 78.00 & 15.26 &  46.72 &  720.3 &  20.70 & 100 \\
A2033 &      &       & Bound-Incoming\tablenotemark{a} & 4.27 & 48.16 &  4.75 &   6.66 &  841.5 &   5.52 & \nodata \\
A2033 &      &       & Bound-Incoming\tablenotemark{a} & 4.60 & 25.94 &  3.53 &   6.33 & 1214.6 &   2.84 & \nodata \\
\hline{}
  SIG & 2.49 & 484.9 & Bound-Outgoing                  & 1.80 & 78.05 & 12.02 &  19.66 &  495.5 &  23.71 &  5 \\
  SIG &      &       & Bound-Incoming                  & 4.70 & 24.39 &  2.73 &   5.40 & 1171.8 &   2.28 & 68 \\
  SIG &      &       & Bound-Incoming                  & 4.08 & 58.59 &  4.78 &   5.99 &  567.7 &   8.23 & 26
\enddata
\tablenotetext{a}{The bound-incoming solutions for the A2029-A2033 system are obtained by assuming the maximum total mass of this system. }
\label{two}
\end{deluxetable*}

\subsection{The Long-Term Future Accretion for the A2029 System}\label{accretion}

The mass contained within the infalling groups constrains the future mass accretion by A2029. 
To place A2029 in the context of other accreting systems, 
 we consider the study by \citet{Haines18} 
 who identify XMM-detected X-ray groups in the infall region of 23 massive clusters at $z \sim 0.2$. 
They calculate the masses of infalling groups using the $M_{200} - L_{X}$ relation \citep{Leauthaud10}. 
The average mass retained in infalling groups per cluster is $2.23 \times 10^{14}$ M$_{\odot}$, 
 corresponding to $\sim19\%$ of the average $M_{200}$ of the primary cluster. 
Because they identify infalling groups within $0.4 \leq R_{cl}/R_{200} \leq 1.3$, 
 \citet{Haines18} use the Millennium Simulation to correct the mass retained in infalling groups outside their survey region. 
They conclude that clusters accrete $16.2 \pm 4.2\%$ of their mass between $z = 0.223$ and the present day. 
Depending primarily on the simulations \citep{Zhao09, Fakhouri10, vandenBosch14}, 
 they also suggest that groups contain only half of the expected total mass accreted by the clusters. 

The sum of masses of the infalling group, A2033 and SIG, is $(6.0 \pm 1.7) \times 10^{14}$ M$_{\odot}$ 
 or $\sim 61 \pm 19\%$ of the A2029 mass, significantly larger than the measurements for the higher redshift clusters. 
If we use the mass estimates of A2029, A2033 and SIG based on the $M_{200} - L_{X}$ relation 
 for direct comparison with \citet{Haines18}, 
 the mass fraction within the infalling group is still large ($\sim 31 \pm 4\%$). 
We note that this mass fraction is a lower limit
 because it does not account for the rest of the mass contained within the infall region.

Because A2033 and SIG lie comfortably within the caustics and 
 because the two-body model has only bound solutions including several inbound trajectories,
 all infalling groups should eventually be accreted onto A2029, 
 implying a high accretion rate compared to the measurements from $z \sim 0.22$ clusters \citep{Haines18}. 
The growth rate also significantly exceeds expectations based on numerical simulations. 
For example, the growth rate of a massive dark matter halo with $10^{15} M_{\odot}$
 derived from the Millennium-II simulation (eq (2) in \citealp{Fakhouri10}) is
 $\sim 1.1 \times 10^{14} M_{\odot}$ ($11\%$ of the DM halo mass)  
 from the A2029 redshift ($z \sim 0.08$, $\sim 1$ Gyr look-back time) to the present day.  
A2029 suggests that stochastic variations in the accretion rate are large.

The caustic method we use for identifying cluster members provides a mass profile of the cluster 
 often extending to the turnaround radius \citep{Diaferio97, Diaferio99}. 
Based on the mass profile from the A2029 caustics, 
 we estimate the mass retained within the entire infall region. 
First, we consider a spherical shell with inner and outer radii of 1.99 Mpc and 3.66 Mpc, respectively; 
 500 kpc smaller than the distance to SIG from A2029 core and 500 kpc larger than the distance to A2033 from A2029 core. 
The mass within this spherical shell is $4.93^{+1.24}_{-1.24} \times 10^{14}$ M$_{\odot}$,
 comparable with the sum of the masses of A2033 and SIG estimated from the $M_{200} - L_{X}$ relation. 
The systematic uncertainty in the masses of A2033 and SIG are large (Table \ref{subprop}): 
 the two systems contain at least half of the mass in this annulus and they probably dominate the mass.
 
The caustic mass profile also provides an estimate of the ultimate halo mass of A2029 \citep{Rines13}. 
Simulations demonstrate that
 most of the mass ($\sim 90\%$) within a radius enclosing an overdensity $\sim 5.6 \rho_{crit}$ is ultimately accreted by the halo
 \citep{Busha05, Dunner06}. 
We refer to this mass as the ultimate mass of the cluster. 
\citet{Rines13} estimate the ultimate masses of 58 clusters ($M_{5.6}$) and 
 demonstrate that the typical ratio between the M$_{200}$ and the ultimate mass is $\sim 1.99$ M$_{200}$.
The ultimate mass of A2029 is $M_{5.6} = 1.58^{+5.04}_{-4.98} \times 10^{15}$ M$_{\odot}$ or ($1.86 \pm 0.36$) M$_{200}$.
This result is completely consistent with \citet{Rines13}.
In summary the current A2029 contain $54\%$ of the ultimate halo mass, 
 A2033 and SIG contain $34 \pm 12\%$ of the ultimate halo mass, 
 and the remainder is distributed throughout the infall region  possibly in lower mass groups (mostly at radii larger than $R_{cl} > 3.7$ Mpc). 

The large mass accretion rate of A2029 is interesting,
 but may not be surprising because it is one of the most massive clusters in the nearby universe. 
The accretion rate as a function of cluster mass and redshift 
 is a powerful constraint on the hierarchical growth of these systems. 
Even with the extensive dataset for A2029, 
 the uncertainties in the dynamical future of the system remain large.
Having a comprehensive observational view of the system extending 
 throughout the infall region is crucial for estimating the ultimate mass of the system.
All the data taken together suggest that A2029 experienced an accretion event 3 Gyr ago and
 will experience one or more events within the next 3 Gyr.

\section{SUMMARY}

We combine a dense redshift survey of the local massive cluster A2029 with X-ray and weak lensing maps
 to elucidate the future accretion story of this massive system. 
The total dataset for A2029 is unusually rich.
The redshift survey is essentially complete within a wide field of $R_{cl} < 40\arcmin (= 3.5$ Mpc) around A2029. 
We refine analysis of the {\it ROSAT} images and of the weak lensing map 
 to improve mass estimated for two massive subsystems, A2033 and SIG, within the A2029 infall region.  

The infalling groups, A2033 and SIG, appear in the weak lensing map and the X-ray image and the spectroscopic survey. 
Interestingly, 
 the brightest galaxies in these subgroups are offset from the group centers 
 (determined by X-ray or cluster members). 
The astrophysical implications of these offsets are unclear. 

The complete redshift survey facilitates the identification of foreground and background groups in the A2029 field. 
This identification is critical for removing spurious contributions to the mass within the infall region.
We identify at least 12 foreground/background systems. 
Among these systems, 10 systems have {\it ROSAT} X-ray counterparts; 
 a very bright X-ray group LOS7 lies $z = 0.223$. 
Oddly its position makes it appear to be a filamentary connection between A2033 and A2029. 
The redshift survey makes it clear that this apparent connection is merely a superposition. 
Taking these extended X-ray sources together with A2029, A2033, and SIG we demonstrate that
 they are all consistent with the well-known scaling relation between X-ray luminosity and velocity dispersion.  

We measure the mass of A2029 based on the three different mass proxies: 
 caustics, weak lensing and X-ray luminosity (or temperature). 
The caustic mass based on the spectroscopic members is $M_{200} = (8.47 \pm 0.25) \times 10^{14} M_{\odot}$, 
 agrees to within $1\sigma$ with the X-ray estimate. 
We also estimate the masses of infalling groups using velocity dispersions, weak lensing and X-ray luminosities. 
Within the much larger uncertainties, the estimates agree. 
They imply that the total mass in these two subsystems in $\sim 60\%$ of the mass of the main cluster.

A simple two-body model traces the future accretion of the infalling groups. 
The model suggests that the infalling groups are obviously bound to A2029 
 and may be accreted by the primary cluster within $\sim 3$ Gyr. 
This accretion rate is larger than the average predicted by simulations. 

The infall region as a whole contains an amount of mass comparable with the A2029 M$_{200}$. 
The two massive subsystems contribute about $\sim 60\%$ of the mass in the infall region. 
Numerical simulations suggest that $90\%$ of the mass in the infall region will be accreted in the long-term future of the cluster.

In the future a combination of eROSITA, PFS, and Euclid observations will make similar analyses
 possible for clusters across a broad range of cluster mass and over a wide redshift range. 
These combined spectroscopic, X-ray and weak lensing observations will enable 
 construction of the full picture of the accretion story of clusters of galaxies. 
They will provide a strong test of the hierarchical structure formation picture.

\acknowledgments
We thank Jacqueline McCleary for her help in providing the code to estimate the lensing masses.
We thank Perry Berlind and Michael Calkins for operating Hectospec and 
 Susan Tokarz for helping with the data reduction. 
This paper uses data products produced by the OIR Telescope Data Center, 
 supported by the Smithsonian Astrophysical Observatory.   
We also thank Gerrit Schellenberger	and Heng Yu for their help for X-ray data analyses. 
J.S. gratefully acknowledges the support of a CfA Fellowship. 
The Smithsonian Institution supported the research of M.J.G.
S.A.W. was supported by an appointment to the NASA Postdoctoral Program at the Goddard Space Flight Center, 
 administered by the Universities Space Research Association through a contract with NASA.
AD acknowledges partial support from 
 the INFN Grant InDark and the grant of the Italian Ministry of Education, 
 University and Research (MIUR) (L. 232/2016) 
 ``ECCELLENZA1822 D206 -Dipartimento di Eccellenza 2018-2022 Fisica" 
 awarded to the Dept. of Physics of the University of Torino. 
This research has made use of NASAs Astrophysics Data System Bibliographic Services. 

Funding for SDSS-III has been provided by the Alfred P. Sloan Foundation, 
 the Participating Institutions, the National Science Foundation, 
 and the U.S. Department of Energy Office of Science. 
The SDSS-III web site is http://www.sdss3.org/. 
SDSS-III is managed by the Astrophysical Research Consortium for 
 the Participating Institutions of the SDSS-III Collaboration including 
 the University of Arizona, the Brazilian Participation Group, 
 Brookhaven National Laboratory, University of Cambridge, 
 Carnegie Mellon University, University of Florida, the French Participation Group, 
 the German Participation Group, Harvard University, the Instituto de Astrofisica de Canarias, 
 the Michigan State/Notre Dame/ JINA Participation Group, Johns Hopkins University, 
 Lawrence Berkeley National Laboratory, Max Planck Institute for Astrophysics, 
 Max Planck Institute for Extraterrestrial Physics, New Mexico State University, 
 New York University, Ohio State University, Pennsylvania State University, 
 University of Portsmouth, Princeton University, the Spanish Participation Group, 
 University of Tokyo, University of Utah, Vanderbilt University, University of Virginia, 
 University of Washington, and Yale University.

\bibliographystyle{apj}
\bibliography{ms}

\begin{thebibliography}{}
\expandafter\ifx\csname natexlab\endcsname\relax\def\natexlab#1{#1}\fi

\bibitem[{{Abell} {et~al.}(1989){Abell}, {Corwin}, \& {Olowin}}]{Abell89}
{Abell}, G.~O., {Corwin}, Jr., H.~G., \& {Olowin}, R.~P. 1989, \apjs, 70, 1

\bibitem[{{Alam} {et~al.}(2015){Alam}, {Albareti}, {Allende Prieto}, {Anders},
  {Anderson}, {Anderton}, {Andrews}, {Armengaud}, {Aubourg}, {Bailey}, \&
  et~al.}]{Alam15}
{Alam}, S., {Albareti}, F.~D., {Allende Prieto}, C., {et~al.} 2015, \apjs, 219,
  12

\bibitem[{{Amendola} {et~al.}(2018){Amendola}, {Appleby}, {Avgoustidis},
  {Bacon}, {Baker}, {Baldi}, {Bartolo}, {Blanchard}, {Bonvin}, {Borgani},
  {Branchini}, {Burrage}, {Camera}, {Carbone}, {Casarini}, {Cropper}, {de
  Rham}, {Dietrich}, {Di Porto}, {Durrer}, {Ealet}, {Ferreira}, {Finelli},
  {Garc{\'{\i}}a-Bellido}, {Giannantonio}, {Guzzo}, {Heavens}, {Heisenberg},
  {Heymans}, {Hoekstra}, {Hollenstein}, {Holmes}, {Hwang}, {Jahnke},
  {Kitching}, {Koivisto}, {Kunz}, {La Vacca}, {Linder}, {March}, {Marra},
  {Martins}, {Majerotto}, {Markovic}, {Marsh}, {Marulli}, {Massey}, {Mellier},
  {Montanari}, {Mota}, {Nunes}, {Percival}, {Pettorino}, {Porciani},
  {Quercellini}, {Read}, {Rinaldi}, {Sapone}, {Sawicki}, {Scaramella},
  {Skordis}, {Simpson}, {Taylor}, {Thomas}, {Trotta}, {Verde}, {Vernizzi},
  {Vollmer}, {Wang}, {Weller}, \& {Zlosnik}}]{Amendola18}
{Amendola}, L., {Appleby}, S., {Avgoustidis}, A., {et~al.} 2018, Living Reviews
  in Relativity, 21, 2

\bibitem[{{Arnaud} {et~al.}(2005){Arnaud}, {Pointecouteau}, \&
  {Pratt}}]{Arnaud05}
{Arnaud}, M., {Pointecouteau}, E., \& {Pratt}, G.~W. 2005, \aap, 441, 893

\bibitem[{{Beers} {et~al.}(1982){Beers}, {Geller}, \& {Huchra}}]{Beers82}
{Beers}, T.~C., {Geller}, M.~J., \& {Huchra}, J.~P. 1982, \apj, 257, 23

\bibitem[{{Ben{\'{\i}}tez}(2000)}]{Benitez00}
{Ben{\'{\i}}tez}, N. 2000, \apj, 536, 571

\bibitem[{{B{\"o}hringer} {et~al.}(2013){B{\"o}hringer}, {Chon}, {Collins},
  {Guzzo}, {Nowak}, \& {Bobrovskyi}}]{Bohringer13}
{B{\"o}hringer}, H., {Chon}, G., {Collins}, C.~A., {et~al.} 2013, \aap, 555,
  A30

\bibitem[{{Bond} {et~al.}(1996){Bond}, {Kofman}, \& {Pogosyan}}]{Bond96}
{Bond}, J.~R., {Kofman}, L., \& {Pogosyan}, D. 1996, \nat, 380, 603

\bibitem[{{Bower} {et~al.}(1988){Bower}, {Ellis}, \& {Efstathiou}}]{Bower88}
{Bower}, R.~G., {Ellis}, R.~S., \& {Efstathiou}, G. 1988, \mnras, 234, 725

\bibitem[{{Boylan-Kolchin} {et~al.}(2009){Boylan-Kolchin}, {Springel}, {White},
  {Jenkins}, \& {Lemson}}]{BoylanKolchin09}
{Boylan-Kolchin}, M., {Springel}, V., {White}, S.~D.~M., {Jenkins}, A., \&
  {Lemson}, G. 2009, \mnras, 398, 1150

\bibitem[{{Busha} {et~al.}(2005){Busha}, {Evrard}, {Adams}, \&
  {Wechsler}}]{Busha05}
{Busha}, M.~T., {Evrard}, A.~E., {Adams}, F.~C., \& {Wechsler}, R.~H. 2005,
  \mnras, 363, L11

\bibitem[{{Clarke} {et~al.}(2004){Clarke}, {Blanton}, \& {Sarazin}}]{Clarke04}
{Clarke}, T.~E., {Blanton}, E.~L., \& {Sarazin}, C.~L. 2004, \apj, 616, 178

\bibitem[{{Colberg} {et~al.}(2005){Colberg}, {Krughoff}, \&
  {Connolly}}]{Colberg05}
{Colberg}, J.~M., {Krughoff}, K.~S., \& {Connolly}, A.~J. 2005, \mnras, 359,
  272

\bibitem[{{Danese} {et~al.}(1980){Danese}, {de Zotti}, \& {di
  Tullio}}]{Danese80}
{Danese}, L., {de Zotti}, G., \& {di Tullio}, G. 1980, \aap, 82, 322

\bibitem[{{De Boni} {et~al.}(2016){De Boni}, {Serra}, {Diaferio}, {Giocoli}, \&
  {Baldi}}]{deBoni16}
{De Boni}, C., {Serra}, A.~L., {Diaferio}, A., {Giocoli}, C., \& {Baldi}, M.
  2016, \apj, 818, 188

\bibitem[{{Diaferio}(1999)}]{Diaferio99}
{Diaferio}, A. 1999, \mnras, 309, 610

\bibitem[{{Diaferio} \& {Geller}(1997)}]{Diaferio97}
{Diaferio}, A., \& {Geller}, M.~J. 1997, \apj, 481, 633

\bibitem[{{D{\"u}nner} {et~al.}(2006){D{\"u}nner}, {Araya}, {Meza}, \&
  {Reisenegger}}]{Dunner06}
{D{\"u}nner}, R., {Araya}, P.~A., {Meza}, A., \& {Reisenegger}, A. 2006,
  \mnras, 366, 803

\bibitem[{{Ebeling} {et~al.}(1998){Ebeling}, {Edge}, {Bohringer}, {Allen},
  {Crawford}, {Fabian}, {Voges}, \& {Huchra}}]{Ebeling98}
{Ebeling}, H., {Edge}, A.~C., {Bohringer}, H., {et~al.} 1998, \mnras, 301, 881

\bibitem[{{Eckert} {et~al.}(2017){Eckert}, {Ettori}, {Pointecouteau},
  {Molendi}, {Paltani}, \& {Tchernin}}]{Eckert17}
{Eckert}, D., {Ettori}, S., {Pointecouteau}, E., {et~al.} 2017, Astronomische
  Nachrichten, 338, 293

\bibitem[{{Erben} {et~al.}(2001){Erben}, {Van Waerbeke}, {Bertin}, {Mellier},
  \& {Schneider}}]{Erben01}
{Erben}, T., {Van Waerbeke}, L., {Bertin}, E., {Mellier}, Y., \& {Schneider},
  P. 2001, \aap, 366, 717

\bibitem[{{Fabricant} {et~al.}(2005){Fabricant}, {Fata}, {Roll}, {Hertz},
  {Caldwell}, {Gauron}, {Geary}, {McLeod}, {Szentgyorgyi}, {Zajac}, {Kurtz},
  {Barberis}, {Bergner}, {Brown}, {Conroy}, {Eng}, {Geller}, {Goddard},
  {Honsa}, {Mueller}, {Mink}, {Ordway}, {Tokarz}, {Woods}, {Wyatt}, {Epps}, \&
  {Dell'Antonio}}]{Fabricant05}
{Fabricant}, D., {Fata}, R., {Roll}, J., {et~al.} 2005, \pasp, 117, 1411

\bibitem[{{Fakhouri} \& {Ma}(2008)}]{Fakhouri08}
{Fakhouri}, O., \& {Ma}, C.-P. 2008, \mnras, 386, 577

\bibitem[{{Fakhouri} {et~al.}(2010){Fakhouri}, {Ma}, \&
  {Boylan-Kolchin}}]{Fakhouri10}
{Fakhouri}, O., {Ma}, C.-P., \& {Boylan-Kolchin}, M. 2010, \mnras, 406, 2267

\bibitem[{{Geller} {et~al.}(1999){Geller}, {Diaferio}, \& {Kurtz}}]{Geller99}
{Geller}, M.~J., {Diaferio}, A., \& {Kurtz}, M.~J. 1999, \apjl, 517, L23

\bibitem[{{Geller} {et~al.}(2013){Geller}, {Diaferio}, {Rines}, \&
  {Serra}}]{Geller13}
{Geller}, M.~J., {Diaferio}, A., {Rines}, K.~J., \& {Serra}, A.~L. 2013, \apj,
  764, 58

\bibitem[{{Giocoli} {et~al.}(2012){Giocoli}, {Tormen}, \& {Sheth}}]{Giocoli12}
{Giocoli}, C., {Tormen}, G., \& {Sheth}, R.~K. 2012, \mnras, 422, 185

\bibitem[{{Gonzalez} {et~al.}(2018){Gonzalez}, {de los Rios}, {Oio}, {Lang},
  {Tagliaferro}, {Dom{\'{\i}}nguez R.}, {Castell{\'o}n}, {Cuevas L.}, \&
  {Valotto}}]{Gonzalez18}
{Gonzalez}, E.~J., {de los Rios}, M., {Oio}, G.~A., {et~al.} 2018, \aap, 611,
  A78

\bibitem[{{Haines} {et~al.}(2018){Haines}, {Finoguenov}, {Smith}, {Babul},
  {Egami}, {Mazzotta}, {Okabe}, {Pereira}, {Bianconi}, {McGee}, {Ziparo},
  {Campusano}, \& {Loyola}}]{Haines18}
{Haines}, C.~P., {Finoguenov}, A., {Smith}, G.~P., {et~al.} 2018, \mnras, 477,
  4931

\bibitem[{{Hao} {et~al.}(2010){Hao}, {McKay}, {Koester}, {Rykoff}, {Rozo},
  {Annis}, {Wechsler}, {Evrard}, {Siegel}, {Becker}, {Busha}, {Gerdes},
  {Johnston}, \& {Sheldon}}]{Hao10}
{Hao}, J., {McKay}, T.~A., {Koester}, B.~P., {et~al.} 2010, \apjs, 191, 254

\bibitem[{{Hoekstra} {et~al.}(2011){Hoekstra}, {Hartlap}, {Hilbert}, \& {van
  Uitert}}]{Hoekstra11}
{Hoekstra}, H., {Hartlap}, J., {Hilbert}, S., \& {van Uitert}, E. 2011, \mnras,
  412, 2095

\bibitem[{{Hwang} {et~al.}(2014){Hwang}, {Geller}, {Diaferio}, {Rines}, \&
  {Zahid}}]{Hwang14}
{Hwang}, H.~S., {Geller}, M.~J., {Diaferio}, A., {Rines}, K.~J., \& {Zahid},
  H.~J. 2014, \apj, 797, 106

\bibitem[{{Kaiser}(1987)}]{Kaiser87}
{Kaiser}, N. 1987, \mnras, 227, 1

\bibitem[{{Kurtz} \& {Mink}(1998)}]{Kurtz98}
{Kurtz}, M.~J., \& {Mink}, D.~J. 1998, \pasp, 110, 934

\bibitem[{{Lauer} {et~al.}(2014){Lauer}, {Postman}, {Strauss}, {Graves}, \&
  {Chisari}}]{Lauer14}
{Lauer}, T.~R., {Postman}, M., {Strauss}, M.~A., {Graves}, G.~J., \& {Chisari},
  N.~E. 2014, \apj, 797, 82

\bibitem[{{Leauthaud} {et~al.}(2010){Leauthaud}, {Finoguenov}, {Kneib},
  {Taylor}, {Massey}, {Rhodes}, {Ilbert}, {Bundy}, {Tinker}, {George}, {Capak},
  {Koekemoer}, {Johnston}, {Zhang}, {Cappelluti}, {Ellis}, {Elvis}, {Giodini},
  {Heymans}, {Le F{\`e}vre}, {Lilly}, {McCracken}, {Mellier},
  {R{\'e}fr{\'e}gier}, {Salvato}, {Scoville}, {Smoot}, {Tanaka}, {Van
  Waerbeke}, \& {Wolk}}]{Leauthaud10}
{Leauthaud}, A., {Finoguenov}, A., {Kneib}, J.-P., {et~al.} 2010, \apj, 709, 97

\bibitem[{{Lewis} {et~al.}(2002){Lewis}, {Stocke}, \& {Buote}}]{Lewis02}
{Lewis}, A.~D., {Stocke}, J.~T., \& {Buote}, D.~A. 2002, \apjl, 573, L13

\bibitem[{{Liu} {et~al.}(2018){Liu}, {Yu}, {Diaferio}, {Tozzi}, {Hwang},
  {Umetsu}, {Okabe}, \& {Yang}}]{Liu18}
{Liu}, A., {Yu}, H., {Diaferio}, A., {et~al.} 2018, ArXiv e-prints,
  arXiv:1806.10864

\bibitem[{{Martinet} {et~al.}(2016){Martinet}, {Clowe}, {Durret}, {Adami},
  {Acebr{\'o}n}, {Hernandez-Garc{\'{\i}}a}, {M{\'a}rquez}, {Guennou}, {Sarron},
  \& {Ulmer}}]{Martinet16}
{Martinet}, N., {Clowe}, D., {Durret}, F., {et~al.} 2016, \aap, 590, A69

\bibitem[{{McBride} {et~al.}(2009){McBride}, {Fakhouri}, \& {Ma}}]{McBride09}
{McBride}, J., {Fakhouri}, O., \& {Ma}, C.-P. 2009, \mnras, 398, 1858

\bibitem[{{McCleary} \& {Dell'Antonio}(2018)}]{McCleary18}
{McCleary}, J., \& {Dell'Antonio}. 2018, \apj, arXiv:in preparation

\bibitem[{{McCleary} {et~al.}(2015){McCleary}, {dell'Antonio}, \&
  {Huwe}}]{McCleary15}
{McCleary}, J., {dell'Antonio}, I., \& {Huwe}, P. 2015, \apj, 805, 40

\bibitem[{{Merloni} {et~al.}(2012){Merloni}, {Predehl}, {Becker},
  {B{\"o}hringer}, {Boller}, {Brunner}, {Brusa}, {Dennerl}, {Freyberg},
  {Friedrich}, {Georgakakis}, {Haberl}, {Hasinger}, {Meidinger}, {Mohr},
  {Nandra}, {Rau}, {Reiprich}, {Robrade}, {Salvato}, {Santangelo}, {Sasaki},
  {Schwope}, {Wilms}, \& {German eROSITA Consortium}}]{Merloni12}
{Merloni}, A., {Predehl}, P., {Becker}, W., {et~al.} 2012, ArXiv e-prints,
  arXiv:1209.3114

\bibitem[{{Neto} {et~al.}(2007){Neto}, {Gao}, {Bett}, {Cole}, {Navarro},
  {Frenk}, {White}, {Springel}, \& {Jenkins}}]{Neto07}
{Neto}, A.~F., {Gao}, L., {Bett}, P., {et~al.} 2007, \mnras, 381, 1450

\bibitem[{{Okabe} {et~al.}(2010){Okabe}, {Okura}, \& {Futamase}}]{Okabe10}
{Okabe}, N., {Okura}, Y., \& {Futamase}, T. 2010, \apj, 713, 291

\bibitem[{{Patel} {et~al.}(2006){Patel}, {Maddox}, {Pearce},
  {Arag{\'o}n-Salamanca}, \& {Conway}}]{Patel06}
{Patel}, P., {Maddox}, S., {Pearce}, F.~R., {Arag{\'o}n-Salamanca}, A., \&
  {Conway}, E. 2006, \mnras, 370, 851

\bibitem[{{Paterno-Mahler} {et~al.}(2013){Paterno-Mahler}, {Blanton},
  {Randall}, \& {Clarke}}]{PaternoMahler13}
{Paterno-Mahler}, R., {Blanton}, E.~L., {Randall}, S.~W., \& {Clarke}, T.~E.
  2013, \apj, 773, 114

\bibitem[{{Regos} \& {Geller}(1989)}]{Regos89}
{Regos}, E., \& {Geller}, M.~J. 1989, \aj, 98, 755

\bibitem[{{Reisenegger} {et~al.}(2000){Reisenegger}, {Quintana}, {Carrasco}, \&
  {Maze}}]{Reisenegger00}
{Reisenegger}, A., {Quintana}, H., {Carrasco}, E.~R., \& {Maze}, J. 2000, \aj,
  120, 523

\bibitem[{{Rines} \& {Diaferio}(2006)}]{Rines06}
{Rines}, K., \& {Diaferio}, A. 2006, \aj, 132, 1275

\bibitem[{{Rines} {et~al.}(2013){Rines}, {Geller}, {Diaferio}, \&
  {Kurtz}}]{Rines13}
{Rines}, K., {Geller}, M.~J., {Diaferio}, A., \& {Kurtz}, M.~J. 2013, \apj,
  767, 15

\bibitem[{{Rines} {et~al.}(2002){Rines}, {Geller}, {Diaferio}, {Mahdavi},
  {Mohr}, \& {Wegner}}]{Rines02}
{Rines}, K., {Geller}, M.~J., {Diaferio}, A., {et~al.} 2002, \aj, 124, 1266

\bibitem[{{Rosat}(2000)}]{Rosat00}
{Rosat}, C. 2000, VizieR Online Data Catalog, 9030

\bibitem[{{Sarazin} {et~al.}(1998){Sarazin}, {Wise}, \&
  {Markevitch}}]{Sarazin98}
{Sarazin}, C.~L., {Wise}, M.~W., \& {Markevitch}, M.~L. 1998, \apj, 498, 606

\bibitem[{{Schirmer} {et~al.}(2004){Schirmer}, {Erben}, {Schneider}, {Wolf}, \&
  {Meisenheimer}}]{Schirmer04}
{Schirmer}, M., {Erben}, T., {Schneider}, P., {Wolf}, C., \& {Meisenheimer}, K.
  2004, \aap, 420, 75

\bibitem[{{Schneider}(1996)}]{Schneider96}
{Schneider}, P. 1996, \mnras, 283, 837

\bibitem[{{Serra} \& {Diaferio}(2013)}]{Serra13}
{Serra}, A.~L., \& {Diaferio}, A. 2013, \apj, 768, 116

\bibitem[{{Serra} {et~al.}(2011){Serra}, {Diaferio}, {Murante}, \&
  {Borgani}}]{Serra11}
{Serra}, A.~L., {Diaferio}, A., {Murante}, G., \& {Borgani}, S. 2011, \mnras,
  412, 800

\bibitem[{{Sif{\'o}n} {et~al.}(2015){Sif{\'o}n}, {Hoekstra}, {Cacciato},
  {Viola}, {K{\"o}hlinger}, {van der Burg}, {Sand}, \& {Graham}}]{Sifon15}
{Sif{\'o}n}, C., {Hoekstra}, H., {Cacciato}, M., {et~al.} 2015, \aap, 575, A48

\bibitem[{{Smith} {et~al.}(2001){Smith}, {Brickhouse}, {Liedahl}, \&
  {Raymond}}]{Smith01}
{Smith}, R.~K., {Brickhouse}, N.~S., {Liedahl}, D.~A., \& {Raymond}, J.~C.
  2001, \apjl, 556, L91

\bibitem[{{Snowden} {et~al.}(1994){Snowden}, {McCammon}, {Burrows}, \&
  {Mendenhall}}]{Snowden94}
{Snowden}, S.~L., {McCammon}, D., {Burrows}, D.~N., \& {Mendenhall}, J.~A.
  1994, \apj, 424, 714

\bibitem[{{Sohn} {et~al.}(2017){Sohn}, {Geller}, {Zahid}, {Fabricant},
  {Diaferio}, \& {Rines}}]{Sohn17a}
{Sohn}, J., {Geller}, M.~J., {Zahid}, H.~J., {et~al.} 2017, \apjs, 229, 20

\bibitem[{{Sohn} {et~al.}(2015){Sohn}, {Hwang}, {Geller}, {Diaferio}, {Rines},
  {Lee}, \& {Lee}}]{Sohn15}
{Sohn}, J., {Hwang}, H.~S., {Geller}, M.~J., {et~al.} 2015, Journal of Korean
  Astronomical Society, 48, 381

\bibitem[{{Szabo} {et~al.}(2011){Szabo}, {Pierpaoli}, {Dong}, {Pipino}, \&
  {Gunn}}]{Szabo11}
{Szabo}, T., {Pierpaoli}, E., {Dong}, F., {Pipino}, A., \& {Gunn}, J. 2011,
  \apj, 736, 21

\bibitem[{{Takada} {et~al.}(2014){Takada}, {Ellis}, {Chiba}, {Greene},
  {Aihara}, {Arimoto}, {Bundy}, {Cohen}, {Dor{\'e}}, {Graves}, {Gunn},
  {Heckman}, {Hirata}, {Ho}, {Kneib}, {Le F{\`e}vre}, {Lin}, {More},
  {Murayama}, {Nagao}, {Ouchi}, {Seiffert}, {Silverman}, {Sodr{\'e}},
  {Spergel}, {Strauss}, {Sugai}, {Suto}, {Takami}, \& {Wyse}}]{Takada14}
{Takada}, M., {Ellis}, R.~S., {Chiba}, M., {et~al.} 2014, \pasj, 66, R1

\bibitem[{{Tyler} {et~al.}(2013){Tyler}, {Rieke}, \& {Bai}}]{Tyler13}
{Tyler}, K.~D., {Rieke}, G.~H., \& {Bai}, L. 2013, \apj, 773, 86

\bibitem[{{Umetsu} \& {Diemer}(2017)}]{Umetsu17}
{Umetsu}, K., \& {Diemer}, B. 2017, \apj, 836, 231

\bibitem[{{Uson} {et~al.}(1991){Uson}, {Boughn}, \& {Kuhn}}]{Uson91}
{Uson}, J.~M., {Boughn}, S.~P., \& {Kuhn}, J.~R. 1991, \apj, 369, 46

\bibitem[{{van den Bosch}(2002)}]{vandenBosch02}
{van den Bosch}, F.~C. 2002, \mnras, 331, 98

\bibitem[{{van den Bosch} {et~al.}(2014){van den Bosch}, {Jiang}, {Hearin},
  {Campbell}, {Watson}, \& {Padmanabhan}}]{vandenBosch14}
{van den Bosch}, F.~C., {Jiang}, F., {Hearin}, A., {et~al.} 2014, \mnras, 445,
  1713

\bibitem[{{von der Linden} {et~al.}(2014){von der Linden}, {Allen},
  {Applegate}, {Kelly}, {Allen}, {Ebeling}, {Burchat}, {Burke}, {Donovan},
  {Morris}, {Blandford}, {Erben}, \& {Mantz}}]{vonderLinden14}
{von der Linden}, A., {Allen}, M.~T., {Applegate}, D.~E., {et~al.} 2014,
  \mnras, 439, 2

\bibitem[{{Walker} {et~al.}(2012){Walker}, {Fabian}, {Sanders}, {George}, \&
  {Tawara}}]{Walker12}
{Walker}, S.~A., {Fabian}, A.~C., {Sanders}, J.~S., {George}, M.~R., \&
  {Tawara}, Y. 2012, \mnras, 422, 3503

\bibitem[{{Wen} {et~al.}(2009){Wen}, {Han}, \& {Liu}}]{Wen09}
{Wen}, Z.~L., {Han}, J.~L., \& {Liu}, F.~S. 2009, \apjs, 183, 197

\bibitem[{{Wen} {et~al.}(2012){Wen}, {Han}, \& {Liu}}]{Wen12}
---. 2012, \apjs, 199, 34

\bibitem[{{Yu} {et~al.}(2016){Yu}, {Diaferio}, {Agulli}, {Aguerri}, \&
  {Tozzi}}]{Yu16}
{Yu}, H., {Diaferio}, A., {Agulli}, I., {Aguerri}, J.~A.~L., \& {Tozzi}, P.
  2016, \apj, 831, 156

\bibitem[{{Yu} {et~al.}(2018){Yu}, {Diaferio}, {Serra}, \& {Baldi}}]{Yu18}
{Yu}, H., {Diaferio}, A., {Serra}, A.~L., \& {Baldi}, M. 2018, \apj, 860, 118

\bibitem[{{Yu} {et~al.}(2015){Yu}, {Serra}, {Diaferio}, \& {Baldi}}]{Yu15}
{Yu}, H., {Serra}, A.~L., {Diaferio}, A., \& {Baldi}, M. 2015, \apj, 810, 37

\bibitem[{{Zhang} {et~al.}(2011){Zhang}, {Andernach}, {Caretta}, {Reiprich},
  {B{\"o}hringer}, {Puchwein}, {Sijacki}, \& {Girardi}}]{Zhang11}
{Zhang}, Y.-Y., {Andernach}, H., {Caretta}, C.~A., {et~al.} 2011, \aap, 526,
  A105

\bibitem[{{Zhao} {et~al.}(2009){Zhao}, {Jing}, {Mo}, \& {B{\"o}rner}}]{Zhao09}
{Zhao}, D.~H., {Jing}, Y.~P., {Mo}, H.~J., \& {B{\"o}rner}, G. 2009, \apj, 707,
  354

\bibitem[{{ZuHone} {et~al.}(2010){ZuHone}, {Markevitch}, \&
  {Johnson}}]{ZuHone10}
{ZuHone}, J.~A., {Markevitch}, M., \& {Johnson}, R.~E. 2010, \apj, 717, 908

\end{thebibliography}

\appendix
\section{Spectroscopically Identified Stars in the A2029 Field}\label{app} 

We identify 97 stars in the A2029 field from our spectroscopic survey. 
These objects are either stars with SDSS spectroscopic redshifts 
 or Hectospec targets that were classified as galaxies based on the SDSS algorithm. 
These stars have absolute radial velocities smaller than $500~\kms$. 
For completeness, we provide a catalog of these stars in the A2029 field 
 including SDSS object ID, Right Ascension, Declination, r-band magnitude, redshift (or blueshift) and its uncertainty
 (Table \ref{allstar}). 

\begin{deluxetable}{lcccc}
\tablecolumns{5}
\tabletypesize{\scriptsize}
\tablewidth{0pt}
\setlength{\tabcolsep}{0.04in}
\tablecaption{The Catalog of Stars in the A2029 Field}
\tablehead{
\colhead{SDSS Object ID} & \colhead{R.A} & \colhead{Decl.} & \colhead{$r_{cModel, 0}$} & \colhead{z}}
\startdata
1237658780557836343 & 227.727719 &   5.757971 &  19.99 & $-0.00008 \pm 0.00009$ \\
1237655744020021696 & 227.711226 &   5.716711 &  20.80 & $-0.00065 \pm 0.00016$ \\
1237655744020086825 & 227.778066 &   5.722604 &  15.20 & $-0.00001 \pm 0.00014$ \\
1237655744020086971 & 227.742827 &   5.697524 &  16.40 & $-0.00018 \pm 0.00003$ \\
1237662268074033278 & 227.833153 &   5.788820 &  15.33 & $ 0.00003 \pm 0.00004$ \\
1237662268074033279 & 227.835948 &   5.790469 &  15.28 & $-0.00014 \pm 0.00004$ \\
1237655744020086887 & 227.848296 &   5.668711 &  19.76 & $-0.00011 \pm 0.00014$ \\
1237658780557770986 & 227.669673 &   5.830535 &  16.44 & $-0.00000 \pm 0.00007$ \\
1237662268074034025 & 227.788274 &   5.871173 &  20.60 & $-0.00034 \pm 0.00013$ \\
1237655744020021260 & 227.609299 &   5.665744 &  16.43 & $-0.00007 \pm 0.00004$
\enddata
\label{allstar}
\tablecomments{
The entire table is available in machine-readable form in the online journal. 
Here, a portion is shown for guidance regarding its format. }
\end{deluxetable}
\clearpage
 
\end{document}